\documentclass{sig-alternate-05-2015}

\usepackage{balance}  % to better equalize the last page
\usepackage{graphics} % for EPS, load graphicx instead
\usepackage{graphicx}
\usepackage{caption}
\usepackage{times}    % comment if you want LaTeX's default font
\usepackage{url}      % llt: nicely formatted URLs

\usepackage{subfigure}
\usepackage{epstopdf}
\usepackage{epsfig}
\usepackage{amsmath}
\usepackage{url}
\usepackage{enumitem}

\newcommand{\remove}[1]{}

\newcommand{\specialcell}[2][c]{%for multi-line cells in tables
  \begin{tabular}[#1]{@{}c@{}}#2\end{tabular}}
\newcommand{\specialcellleft}[2][c]{%
  \begin{tabular}[#1]{@{}l@{}}#2\end{tabular}}

\makeatletter
\def\url@leostyle{%
  \@ifundefined{selectfont}{\def\UrlFont{\sf}}{\def\UrlFont{\small\bf\ttfamily}}}
\makeatother
\def\pprw{8.5in}
\def\pprh{11in}

\def\P{\mathbb{P}}

\usepackage[pdftex]{hyperref}
\hypersetup{
pdftitle={iPhone's Digital Marketplace: Characterizing the Big Spenders},
pdfauthor={Farshad Kooti},
pdfkeywords={App purchases; demographics; prediction; iPhone},
bookmarksnumbered,
pdfstartview={FitH},
colorlinks,
citecolor=black,
filecolor=black,
linkcolor=black,
urlcolor=black,
breaklinks=true,
}

\begin{document}

\CopyrightYear{2017}
\setcopyright{acmcopyright}
\conferenceinfo{WSDM 2017,}{February 06-10, 2017, Cambridge, United Kingdom}
\isbn{978-1-4503-4675-7/17/02}\acmPrice{\$15.00}
\doi{http://dx.doi.org/10.1145/3018661.3018697}

\title{iPhone's Digital Marketplace:\\Characterizing the Big Spenders}

\numberofauthors{5}
\author{
\alignauthor Farshad Kooti \\
\affaddr{USC/ISI} \\ \affaddr{Marina del Rey, USA} \\
\email{kooti@usc.edu}
\alignauthor Mihajlo Grbovic \\
\affaddr{Airbnb} \\ \affaddr{San Francisco, USA} \\
\email{mihajlo.grbovic@airbnb.com}
\alignauthor Luca Maria Aiello\\
\affaddr{Bell Labs} \\ \affaddr{Cambridge, UK} \\
\email{luca.aiello@nokia-bell-labs.com}\\
\and
\alignauthor Eric Bax\\
\affaddr{Yahoo Research} \\ \affaddr{Playa Vista, USA} \\
\email{ebax@yahoo-inc.com}
\alignauthor Kristina Lerman\\
\affaddr{USC/ISI} \\ \affaddr{Marina del Rey, USA} \\
\email{lerman@isi.edu}
}

\date{\today}

\maketitle

%TYPICAL NUMBER OF DAYS BEFORE THE PEAK?

\begin{abstract}

With mobile shopping surging in popularity, people are spending ever more money on digital purchases through their mobile devices and phones. However, few large-scale studies of mobile shopping exist. In this paper we analyze a large data set consisting of more than 776M digital purchases made on Apple mobile devices that include songs, apps, and in-app purchases. We find that 61\% of all the spending is on in-app purchases and that the top 1\% of users are responsible for 59\% of all the spending. These big spenders are more likely to be male and older, and less likely to be from the US. We study how they adopt and abandon individual app, and find that, after an initial phase of increased daily spending, users gradually lose interest: the delay between their purchases increases and the spending decreases with a sharp drop toward the end. Finally, we model the in-app purchasing behavior in multiple steps: $1)$ we model the time between purchases; $2)$ we train a classifier to predict whether the user will make a purchase from a new app or continue purchasing from the existing app; and $3)$ based on the outcome of the previous step, we attempt to predict the exact app, new or existing, from which the next purchase will come. The results yield new insights into spending habits in the mobile digital marketplace.

\end{abstract}


\section{Introduction}\label{section:intro}

%~\cite{Malmgren08pnas}

%Online spending
\noindent Consumer spending is an essential component of the economy, accounting for 71\% of the total US gross domestic product (GDP) in 2013~\cite{spending_gdp}. In spite of representing just over 10\% of all consumer spending~\cite{online_small}, online shopping is rapidly growing as people are becoming more comfortable with the online payment systems, security and delivery of the purchased goods. Over the last three years, online sales grew over $45\%$~\cite{bhatnagar00risk} and are showing signs of exponential growth.
%
%While online shopping is still just 10\% of all the spending on goods purchases~\cite{online_small}, it is growing quickly as people are becoming better acquainted with online payment technologies, and online shopping is becoming more convenient. Over the last three years, online sales grew over $45\%$~\cite{bhatnagar00risk}.
%
%App store income / why this study is important
One of the largest and fastest-growing online markets is Apple's iOS market, where people can make digital purchases from several different categories. Apple's revenue from digital purchases surpassed \$20 billion in 2015~\cite{app_high_spending} and more than doubled in just three years~\cite{app_rev1,app_rev2}.

Mobile digital markets offer a wide variety of digital goods, including apps, songs, movies, and digital books. They also include items that users can buy within an app, called {\it in-app purchases}, such as virtual currencies, bonuses, extra game lives and levels, etc. Despite the popularity of the iOS market, there has not been a large-scale study characterizing users' spending on different types of content and apps. For example, it is not known how much is spent on in-app purchases compared to songs. Learning how people spend their money in this context has direct practical implications on the business of several stakeholders, including app developers and managers of online app stores, but it also has important theoretical implications for the understanding of consumer behavior in an emerging market whose dynamics are still poorly known.

%What we study / perks of the data
We study a longitudinal data set extracted from hundreds of millions of email receipts for digital purchases on iPhones, iPads, and other iOS devices (which we refer to as ``iPhone purchases'' for brevity). Besides its large scale, our data has two unprecedented advantages over other data collected in the past. First, besides recording the purchase history, it also includes \textit{demographic information} such as people's age, gender, and country of residence. This enables us not only to characterize the consumer population, but also study how spending differs by income and location. For example, we find that the average spending is not correlated with income but strongly depends on the country of residence. And second, in contrast to the application-centric view of previous studies, our data is user-centric and allows for the observation of user spending behavior across \textit{multiple apps}. This allows us to study how users abandon an app and start using a new one, and to compare the behavior of users who make purchases from a single app with users who make purchases from multiple apps simultaneously.

%Moreover, we also model the user's behavior more accurately and build a model that explains all the purchases of a user. First,
We analyze how users spend money across different categories and show that a small fraction of users are responsible for the majority of spending. We call these users \textit{big spenders}. Moreover, our analysis informs predictive models of digital spending behavior. We first show that the time between purchases is best described by a Pareto distribution. We then build a supervised classifier that accurately predicts whether a user will make a purchase from a new app or one with previous purchases. Finally, based on the outcome of the previous step, we predict the exact app from which the user will purchase.

In summary, our main findings and contributions are as follows:

\begin{itemize}
\item In-app purchases account for 61\% of all money spent on digital purchases on iPhones, followed by songs (23\%) and app purchases (7\%).
\item The spending is highly heterogeneous: the top 1\% of spenders account for 59\% of all money spent on in-app purchases. %The Gini coefficient is 0.884.
\item Big spenders tend to be 3-8 years older, 22\% more likely to be a man, and 31\% less likely to be from the US, compared to the typical spender. Interestingly, income is comparable between the big spenders and the rest of the users.
\item Big spenders become slower to repurchase from an app as time passes, but their rate of spending within an app initially increases, then decreases.
\item From the perspective of app developers and ad networks, big spenders are the most valuable targeting segment. Even if they abandon an app that they are frequently buying from, they are 4.5x more likely to be a big spender in a new app compared to a random user.
\item We model the purchasing process in several steps: modeling time between purchases, predicting purchase from a new app or an app from which the user has already purchased, and predicting the exact app of an in-app purchase.%\note{This isn't clear, it would be better to explicitly state the findings}
\end{itemize}

%\note{check the statements below, if too strong revert to the original version}
Both consumers and producers in the mobile app market might benefit from this study. Our results can inform the deployment of better app recommendation systems so that people download apps they are more likely to enjoy, thus creating higher revenue opportunities for app developers.

%OLD CONCLUSION (ORIGINAL VERSION)
%Our findings can be leveraged for better app recommendations to users and to extend usage of apps by users.

\section{Data Set and Marketplace}\label{section:dataset}

\noindent Shortly after each digital purchase on an iPhone (or any other iOS device), the user receives a confirmation email with details of the purchase. This email contains information about the purchase, including the amount of money spent and the type of purchase. The email has a specific format, making it easy to parse automatically. We obtained information about digital purchases of Yahoo Mail users, using an automated pipeline that hashes the names and IDs of users to preserve anonymity. All the analyses were performed in aggregate and on unidentifiable data. We gathered data covering 15 months, from March 2014 to June 2015. Our data set includes 26M users, who together made more than 776M purchases totaling \$4.6B.

\begin{figure}[t!]
\begin{center}
\includegraphics[width=0.9\columnwidth]{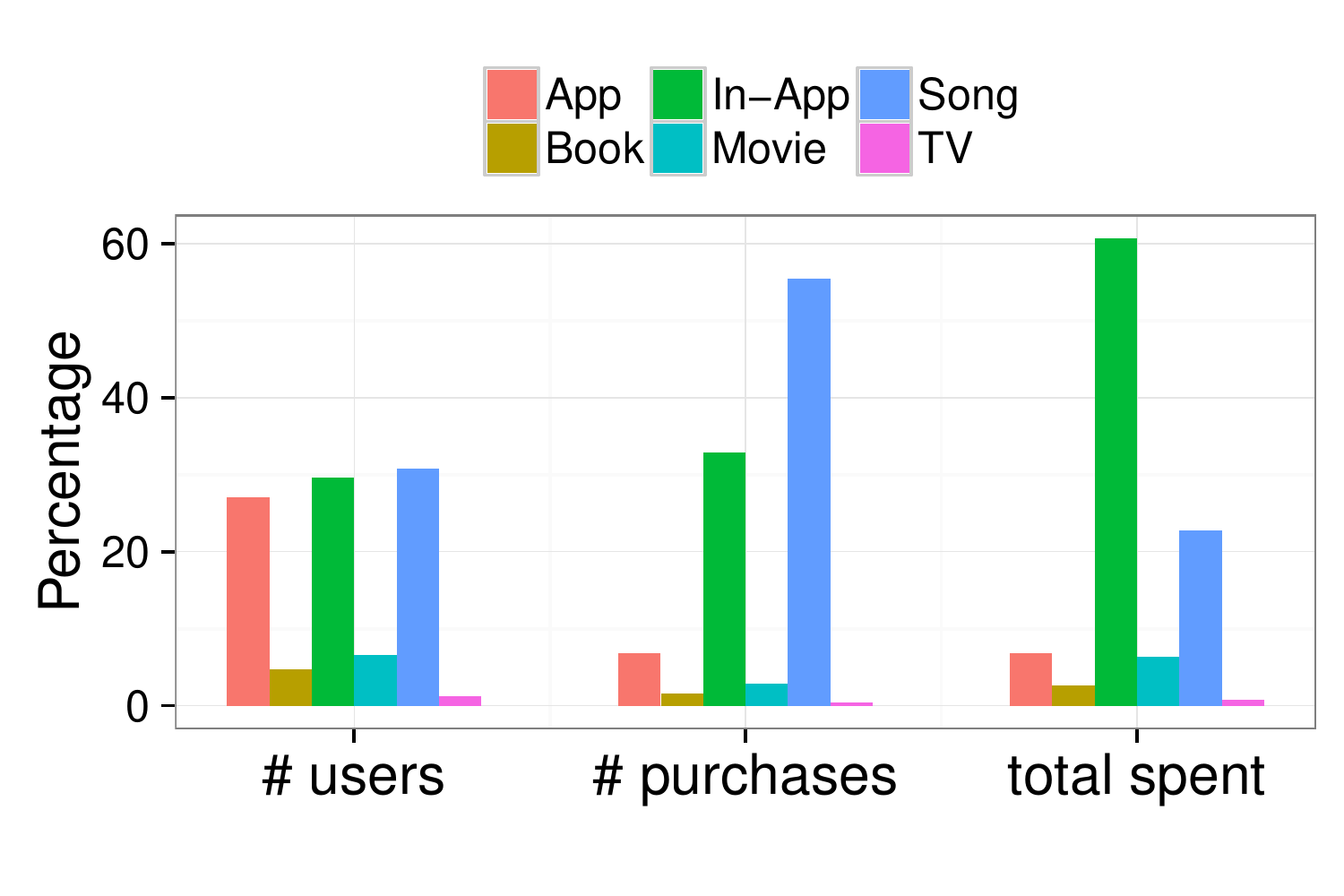}
\end{center}
\vspace*{-1mm}
\caption{Percentage of users, purchases, and money spent on each category.}
\label{fig:bar_all}
\vspace*{-2mm}
\end{figure}

There are six main categories of iPhone purchases: applications (apps), songs, movies, TV shows, books, and in-app purchases (purchases within an app, e.g., bonuses or coins in games). These categories differ vastly in numbers of purchasers: 16M people purchased at least one song, but only 671K people purchased a TV show. The number of purchases by category varies greatly as well: there are 430M song and 255M in-app purchases, while movies, books, and TV shows have fewer than 40M purchases all together. The total money spent in each category varies even more: in-app purchases account for \$2.8B, or 61\%, of all money spent on digital purchases; 23\% of the money is spent on songs, 7\% on app purchases (purchasing apps themselves, not the purchases within apps), 6\% on movies, 2\% on books, and only 0.7\% on TV shows. Even though there are considerably fewer total in-app purchases compared to songs (60\% fewer), the money spent on in-app purchases is 2.7 times more than on songs, showing that on average an in-app purchase is much more expensive than a song purchase. Figure~\ref{fig:bar_all} shows the number of users, purchases, and the money spent in each of these six categories.

\begin{figure}[b!]
\begin{center}
\begin{tabular}{@{}c@{}c@{}}
\subfigure[Spending and gender]{
\includegraphics[width=0.49\columnwidth]{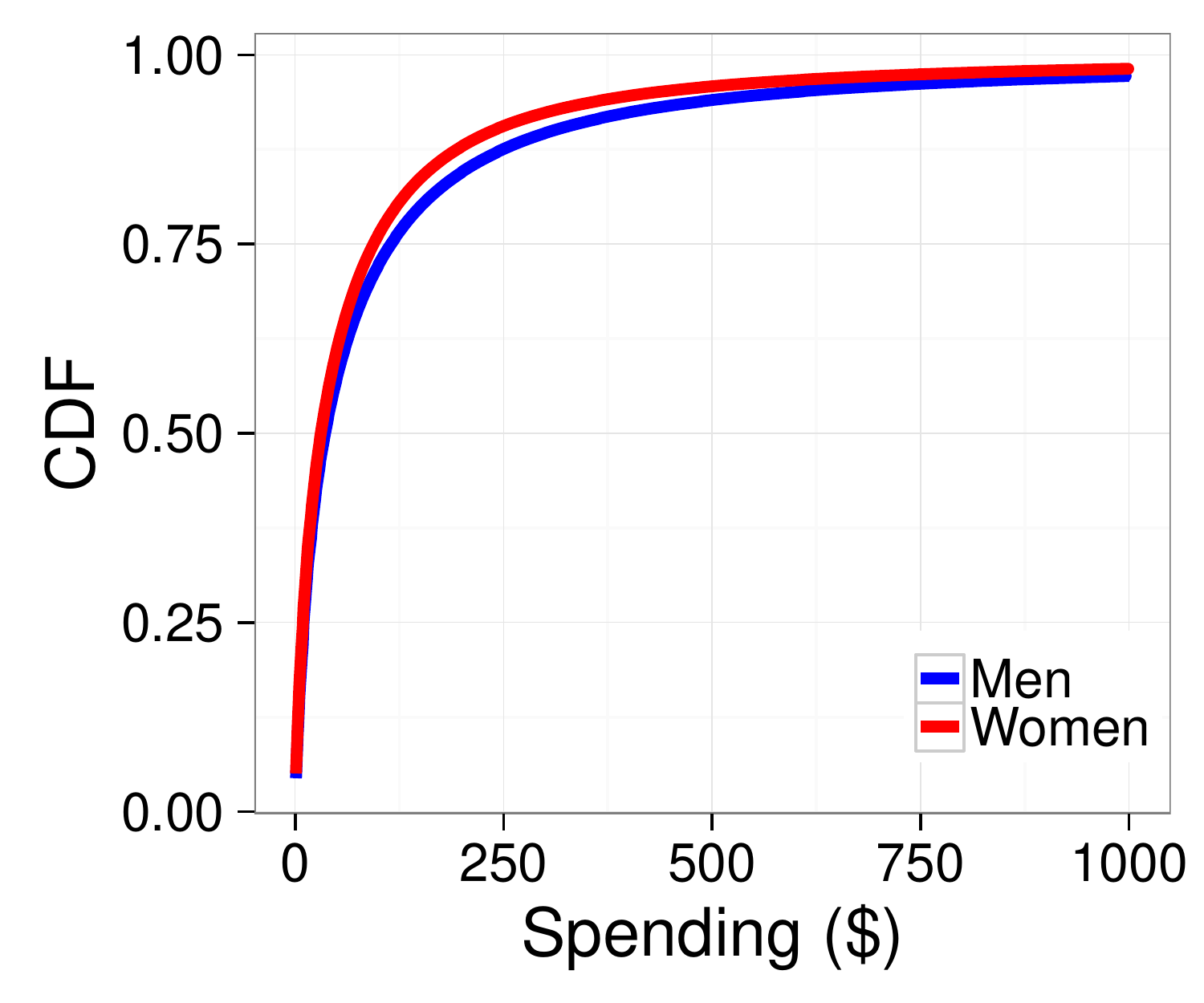}
\label{fig:spending_gender}
}
&
\subfigure[spending and age]{
\includegraphics[width=0.49\columnwidth]{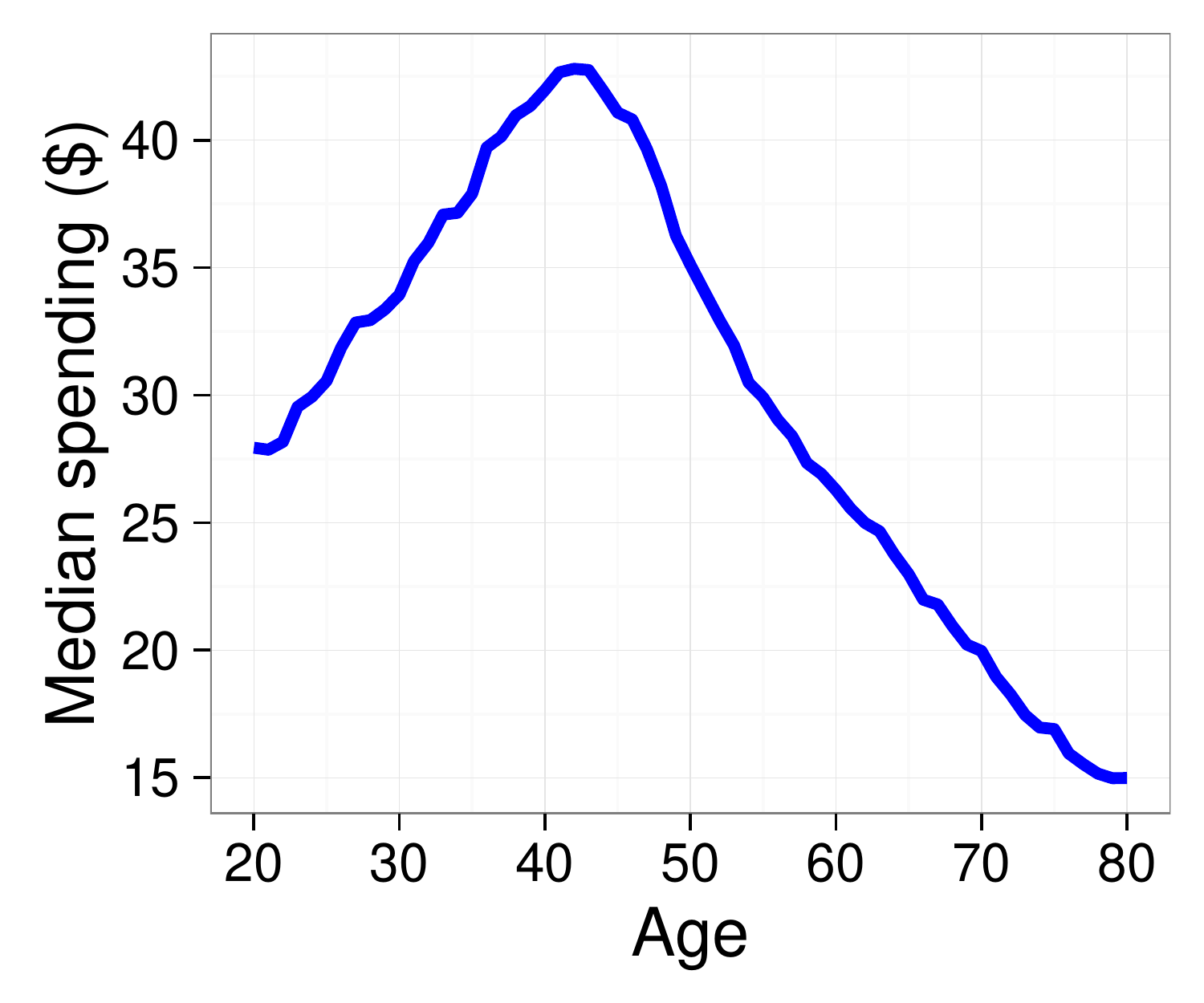}
\label{fig:spending_age}
}
\end{tabular}
\vspace*{-1mm}
\caption{Effect of gender and age on iPhone purchases.}
\label{fig:spending_gender_age}
\end{center}
\end{figure}

\begin{figure}[tbh!]
\begin{center}
\includegraphics[width=0.8\columnwidth]{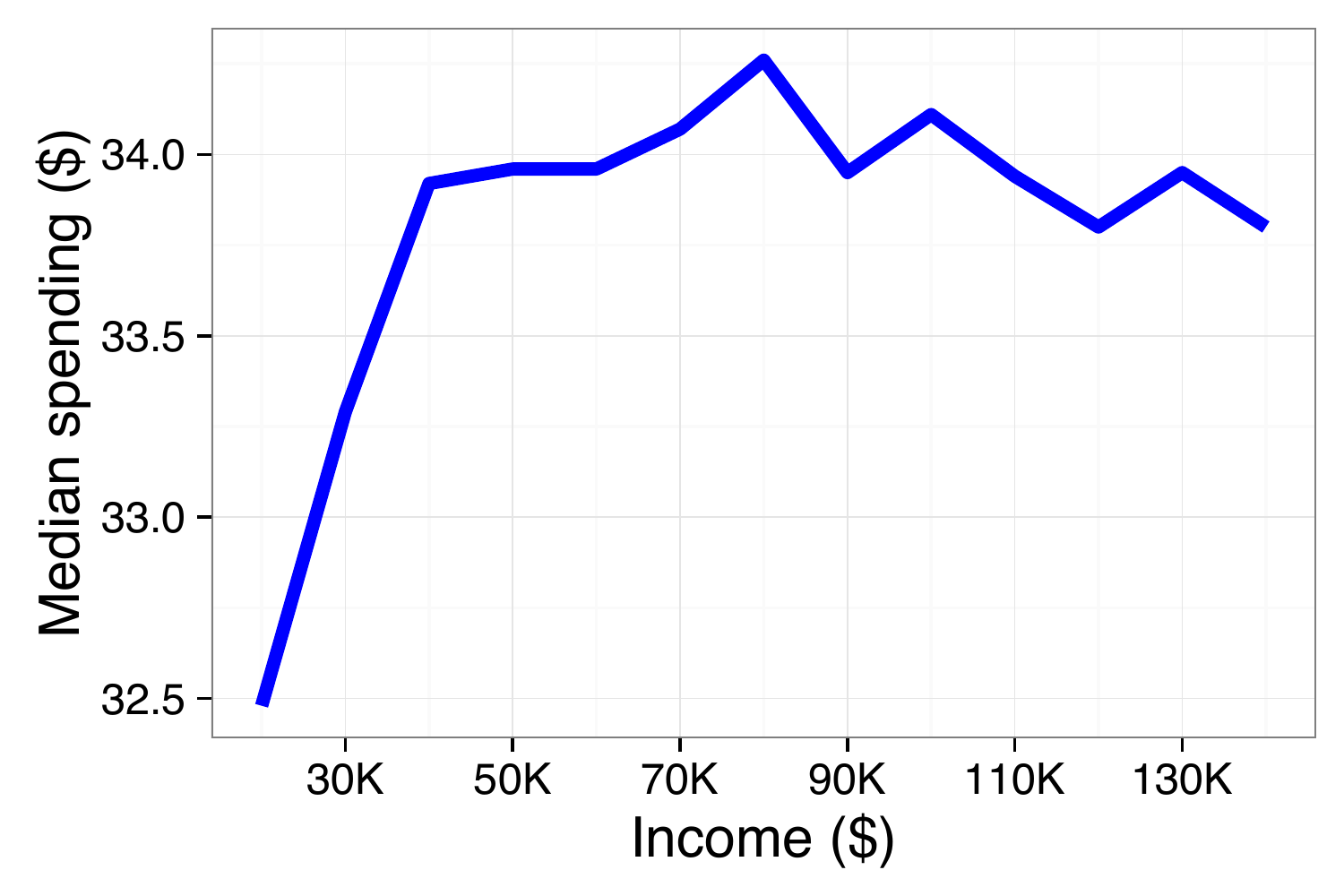}
\end{center}
\caption{Effect of income on spending. There are more than 10k users in each income category.}
\label{fig:spending_income}
\vspace*{-2mm}
\end{figure}

Our data set also includes user age, gender, and zip code as provided by the users at the time of sign up. Spending varies significantly based on demographics. Figure~\ref{fig:spending_gender} shows the cumulative distribution function (CDF) of the spending for men and women. Men spend more money on purchases than women: the median spending for women is \$31.1, and for men it is \$36.2, which is 17\% higher. Age also affects spending: Figure~\ref{fig:spending_age} shows that the peak age for iPhone spenders is the mid-30s, and after that the spending level decreases quickly. For US residents we use the median income of the zip code they declared as their residence as an estimate of their income. This estimate has shown meaningful trends in earlier work~\cite{kooti2015portrait}. Surprisingly, income only affects the spending of people with less than \$40K annual income (Figure~\ref{fig:spending_income}). This is in contrast with online shopping, where users with higher income tend to spend more money shopping online~\cite{kooti2015portrait}.

\begin{figure*}[tbh!]
\begin{center}
\begin{tabular}{@{}c@{}c@{}}
\subfigure[World]{
\includegraphics[width=1.3\columnwidth]{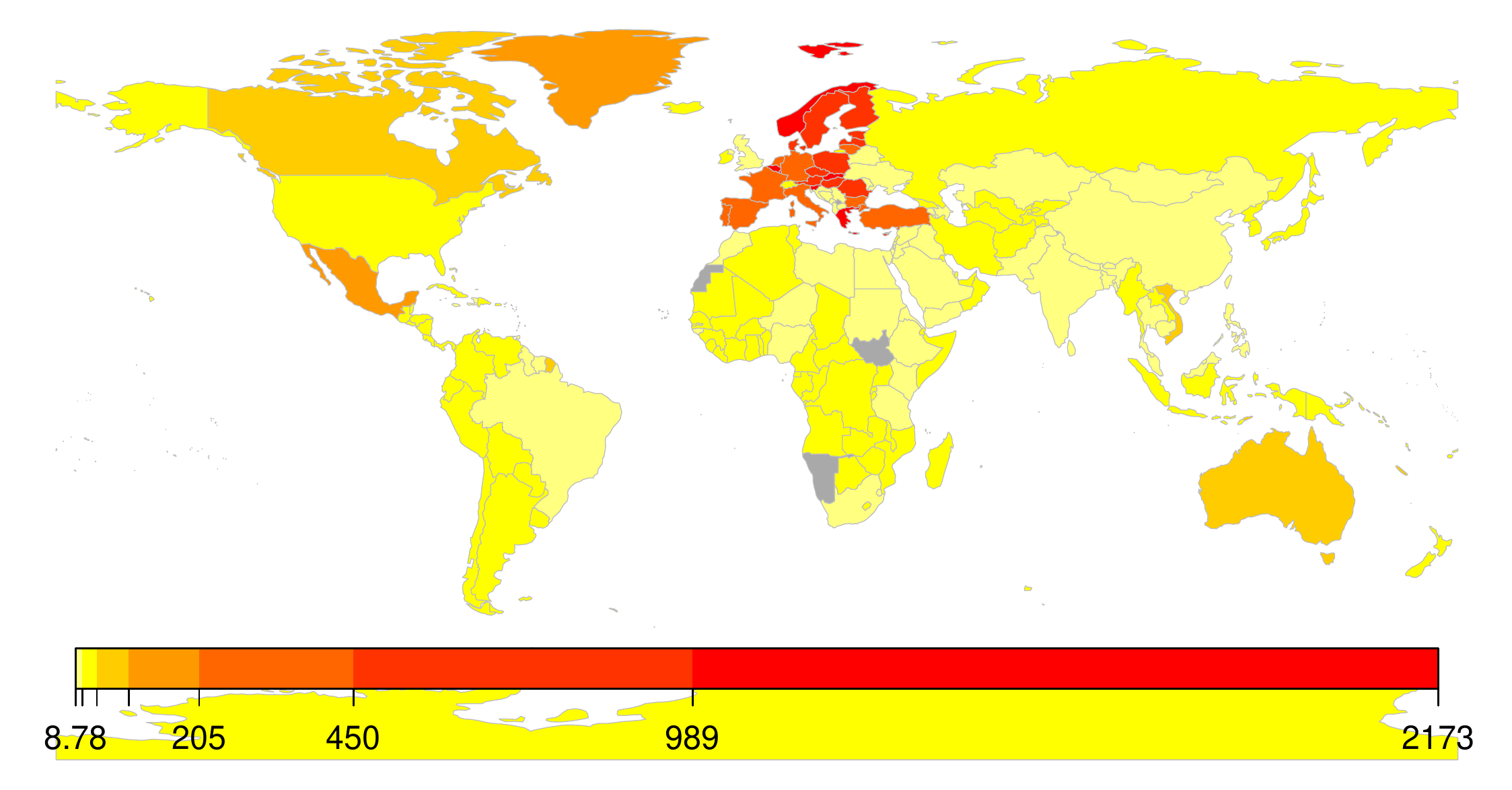}
\label{fig:world_heatmap}
}
&
\subfigure[US]{
\includegraphics[width=0.75\columnwidth]{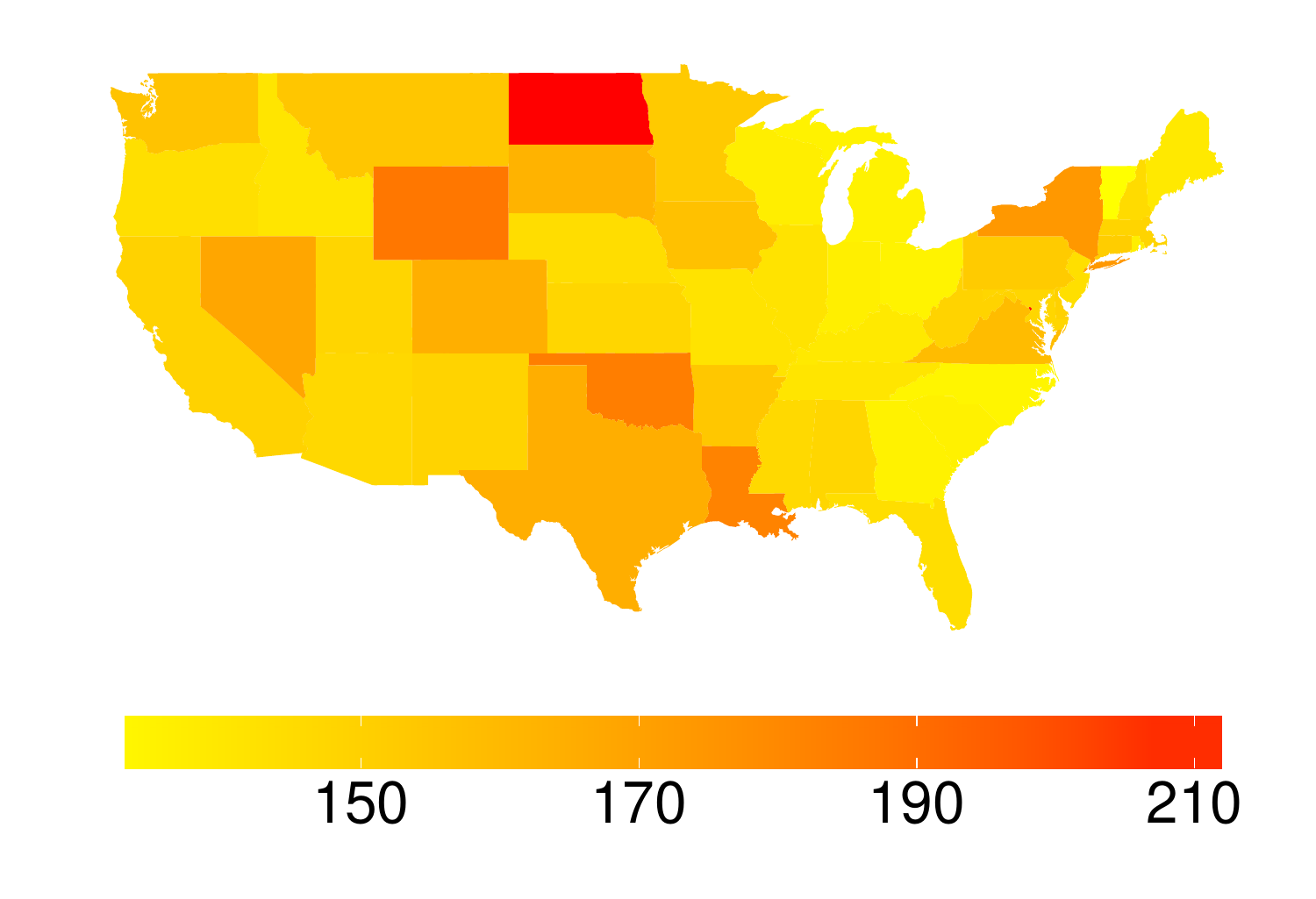}
\label{fig:states_heatmap}
}
\end{tabular}
\vspace*{-2mm}
\caption{Heatmap of the median amount of money spent by the users in each country and US.}
\label{fig:spending_heatmap}
\end{center}
\vspace*{-2mm}
\end{figure*}

Spending on iPhones varies considerably by geography. Figure~\ref{fig:world_heatmap} shows that European countries, especially Scandinavian countries, have the highest spending per Yahoo mail user.  This higher spending could be explained by higher income and fewer people making app purchases. Canada, Mexico, and Australia have higher spending per person than the US, while most African and Asian countries have lower levels of spending.

There are also more than 154K applications in our data set with at least one user purchase or in-app purchase. As shown above the earnings from in-app purchases are considerably higher than from app purchases themselves (almost 9 times higher). Table~\ref{table:top_apps} shows the top 10 apps by in-app earnings, along with their numbers of purchases, average purchaser age, and percentage of female purchasers. We make some observations about these data. First, there are considerable differences among the earnings of top apps. Second, the average price of in-app purchases varies widely across different apps: while there are more than 3.3 times as many purchases in \emph{Candy Crush} compared to Clash of Clans, the earnings for \emph{Clash of Clans} is 2.1 times higher than for \emph{Candy Crush}. Third, there is a significant difference in the demographics of purchasers of different apps. The average age of buyers is 33 years for \emph{Clash of Clans} and 49.4 years for \emph{DoubleDown Casino} and  81\% of the users making purchases from \emph{Farm Heroes Saga} are women, compared to only 18\% for \emph{Boom Beach}. Other apps, such as \emph{Pandora Radio} and \emph{Netflix} (not shown in the table), have a balanced audience. Knowing the audience of an application could be useful for both advertisers and app designers. Advertisers can target the particular population and app designers can make changes to their apps to make them more appealing to the audience that is not currently engaged.

\begin{table}[t!]
\caption{Top 10 apps by in-app earnings, with demographics.}
\begin {center}
{
\scalebox{0.95} {
\begin {tabular} {l | r | r  | c | c }
{App Name} & \multicolumn{1}{|c|}{Earnings} & \multicolumn{1}{|c|}{\specialcell{\# of\\Purchases}} & \specialcell{Avg.\\Age} &  \% Women \\
\hline
\hline
Clash of Clans & \$356.5M & 13.6M & 33.0 & 29\% \\
Candy Crush Saga & \$168.0M & 45.3M & 40.3 & 70\% \\
Game of War & \$159.8M & 1.9M & 35.0 & 25\% \\
Boom Beach & \$60.9M & 1.7M & 34.4 & 18\% \\
Hay Day & \$52.0M & 3.2M & 36.8 & 67\% \\
Farm Heroes Saga & \$40.4M & 6.8M & 42.3 & 81\% \\
Candy Crush Soda& \$37.5M & 8.8M & 41.0 & 77\% \\
Big Fish Casino & \$32.9M & 1.1M & 44.5 & 57\% \\
DoubleDown Casino & \$27.2M & 1.1M & 49.4 & 65\% \\
Pandora Radio & \$24.3M & 5.7M & 37.8 & 54\% \\
\end{tabular}
}
}
\end{center}
\label{table:top_apps}
\vspace*{-3mm}
\end{table}

We also collected the category information for each application, e.g., puzzle game or travel, from Apple's iTunes. Then, for each category we calculated the percentage of people who purchased an app or made an in-app purchase from an app belonging to that category. The top five categories by gender are shown in Table~\ref{table:cat_gender}.  Men are more likely to make a purchase from apps relating to sports, and women prefer games, especially brain games. Similarly, we found the 5 categories with the youngest and oldest average ages.  Younger buyers are interested in photo and video apps, racing games, and social networking applications, while older buyers are interested in more general applications for food, weather, business, and travel (Table~\ref{table:cat_age}).

\begin{table}[t!]
\caption{Top 5 gender-biased categories. }
\begin {center}
{
\scalebox{0.95} {
\begin {tabular} {l | r | l  | r }
\specialcellleft{Top Categories\\for Men} & \% Men & \specialcellleft{Top Categories\\for Women} &  \% Women \\
\hline
\hline
Sport magazines & 84.6\%& Board games & 70.5\%\\
Sports & 74.9\%  & Word games & 64.9\%\\
Racing games &69.9\%  & Puzzle games & 63.6\%\\
Navigation & 68.1\% & Family games & 62.1\%\\
Sports games & 67.4\% & Educational games & 61.5\% \\
\end{tabular}
}
}
\end{center}
\label{table:cat_gender}
\vspace*{-2mm}
\end{table}

\begin{table}[t!]
\caption{Top 5 age-biased categories. }
\begin {center}
{
\scalebox{0.95} {
\begin {tabular} {l | r | l  | r }
\specialcellleft{Categories\\ for Youth} & Avg. Age & \specialcellleft{Categories for\\Older Users} &  Avg. Age \\
\hline
\hline
Photo \& Video & 32.2 & Food \& Drink &47.7\\
Strategy games & 33.2 & Weather &45.4\\
Racing games& 33.6& Travel magazines & 45.2\\
Trivia games& 33.6 & Board games & 44.7\\
Social networking& 34.1 & Business & 42.9 \\
\end{tabular}
}
}
\end{center}
\label{table:cat_age}
\vspace*{-2mm}
\end{table}

{\bf Limitations.} Our data set only includes purchases from people who are Yahoo Mail users; there may be a selection bias in the subset of users being studied. While this might occur to some extent, given the popularity of Yahoo Mail (over 300M users\footnote{\small{\url{www.comscore.com}}}), we believe our data includes a somewhat representative sample and the findings can be generalized to other users. Moreover, the email receipts are sent within a day after the purchase, so our data set does not include the exact time of the purchase, and we have to conduct our analysis with one-day granularity.

\section{Big spenders}\label{section:avid}

\noindent 
In this section we focus on the category of in-app purchases since it is the largest spending category in the digital marketplace. We first show that a small number of users are responsible for the majority of spending. Then we characterize these users demographically, by age, gender, country of residence, and income. Finally, we study how these buyers discover a specific app, start spending money within it, and how they stop making purchases within it as their interest in that app diminishes.

\begin{figure}[tbh!]
\begin{center}
\begin{tabular}{@{}c@{}c@{}}
\subfigure[PDF]{
\includegraphics[width=0.49\columnwidth]{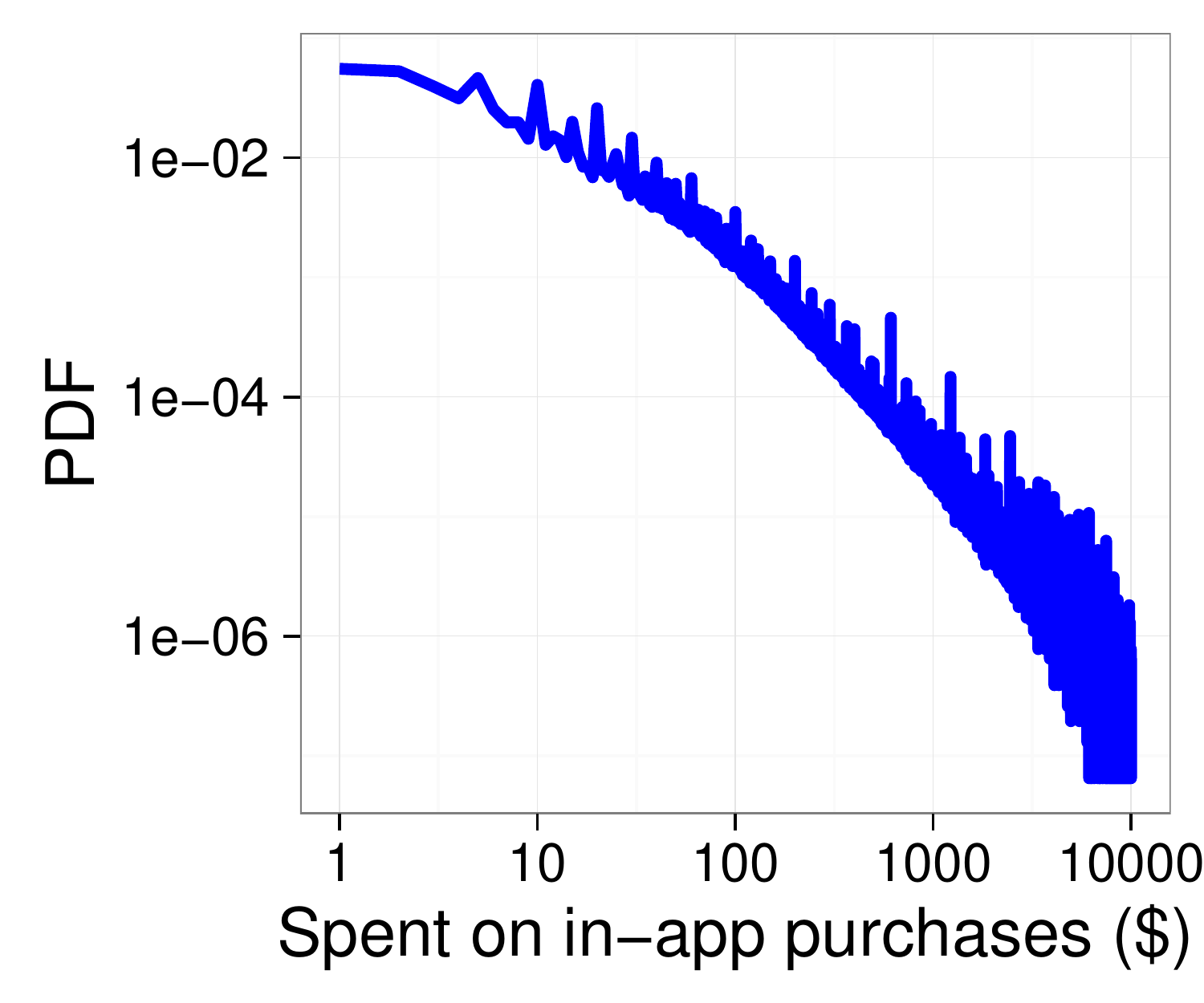}
\label{fig:in_app_apps_pdf}
}
&
\subfigure[CDF]{
\includegraphics[width=0.49\columnwidth]{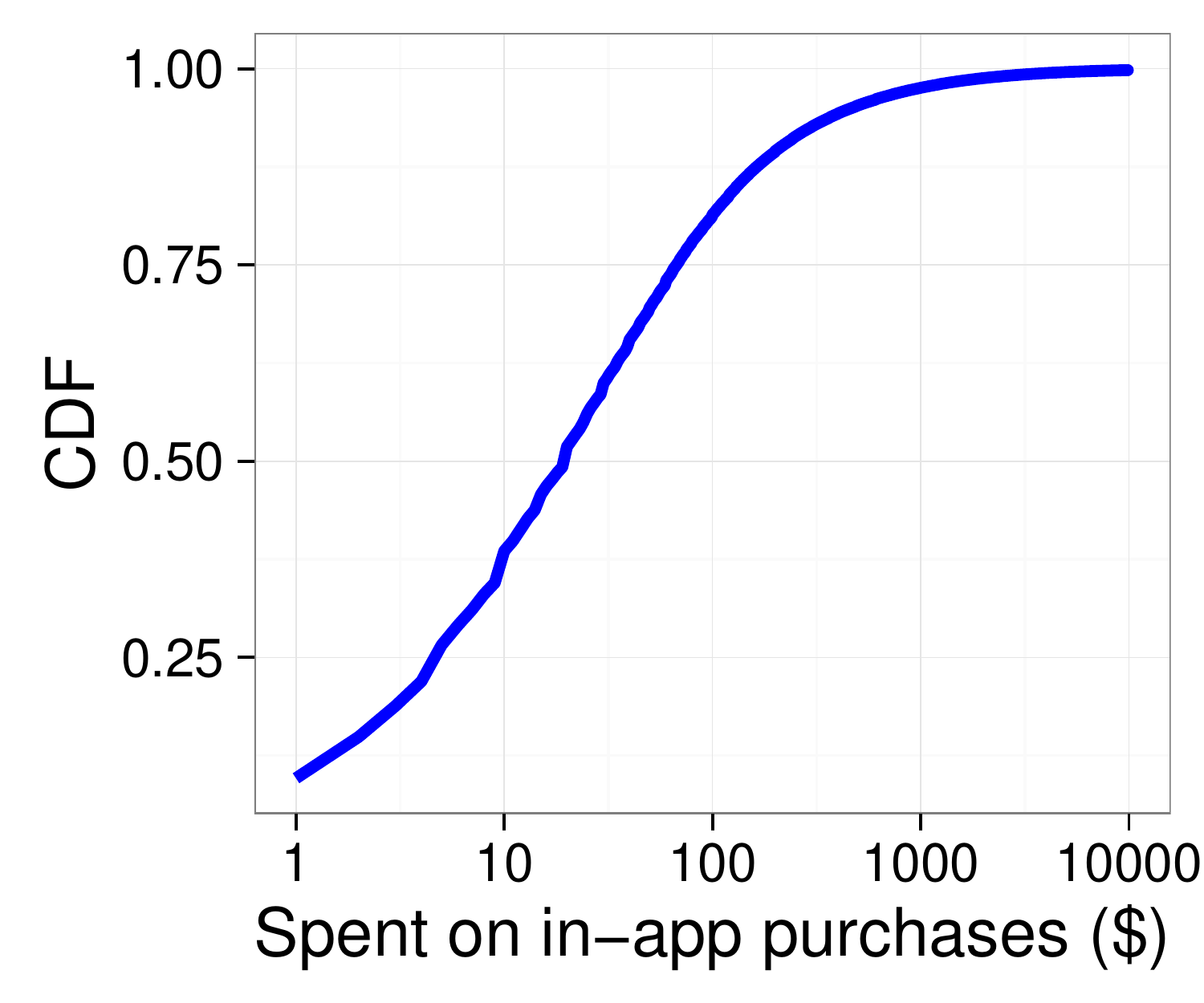}
\label{fig:in_app_apps_cdf}
}
\end{tabular}
\vspace*{-1mm}
\caption{PDF and CDF of user's spending on in-app purchases.}
\label{fig:in_app_apps_pdf_cdf}
\end{center}
\vspace*{-1mm}
\end{figure}

In-app spending patterns vary significantly across different users. In Figure~\ref{fig:in_app_apps_pdf_cdf} we show the PDF and CDF of spending on in-app purchases. It demonstrates that the spending has a heavy-tailed distribution, with most users spending \$20 or less, and a small minority (2.4\% of users) spending more than \$1000 over the studied 15 months.

%over 15 months. However, a small minority spends a large amount of money: 2.4\% of people spent more than \$1000.

\begin{figure}[t!]
\begin{center}
\includegraphics[width=0.7\columnwidth]{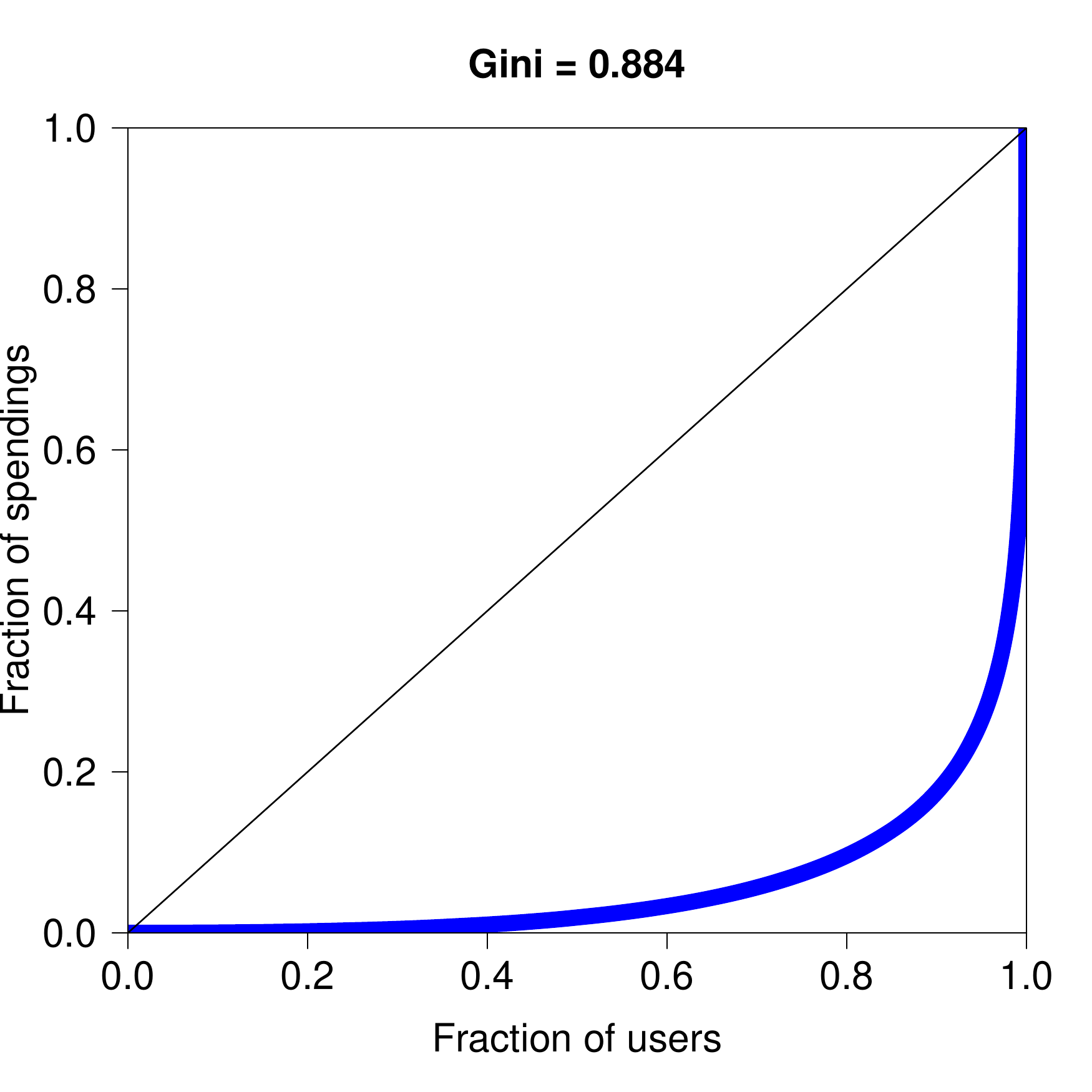}
\end{center}
\vspace*{-1mm}
\caption{Lorenz curve of the spending of the users on in-app purchases, showing high disparity among users.}
\label{fig:lorenz_curve_in_app_spending}
\vspace*{-1mm}
\end{figure}

\begin{figure}[tbh!]
\begin{center}
\includegraphics[width=0.7\columnwidth]{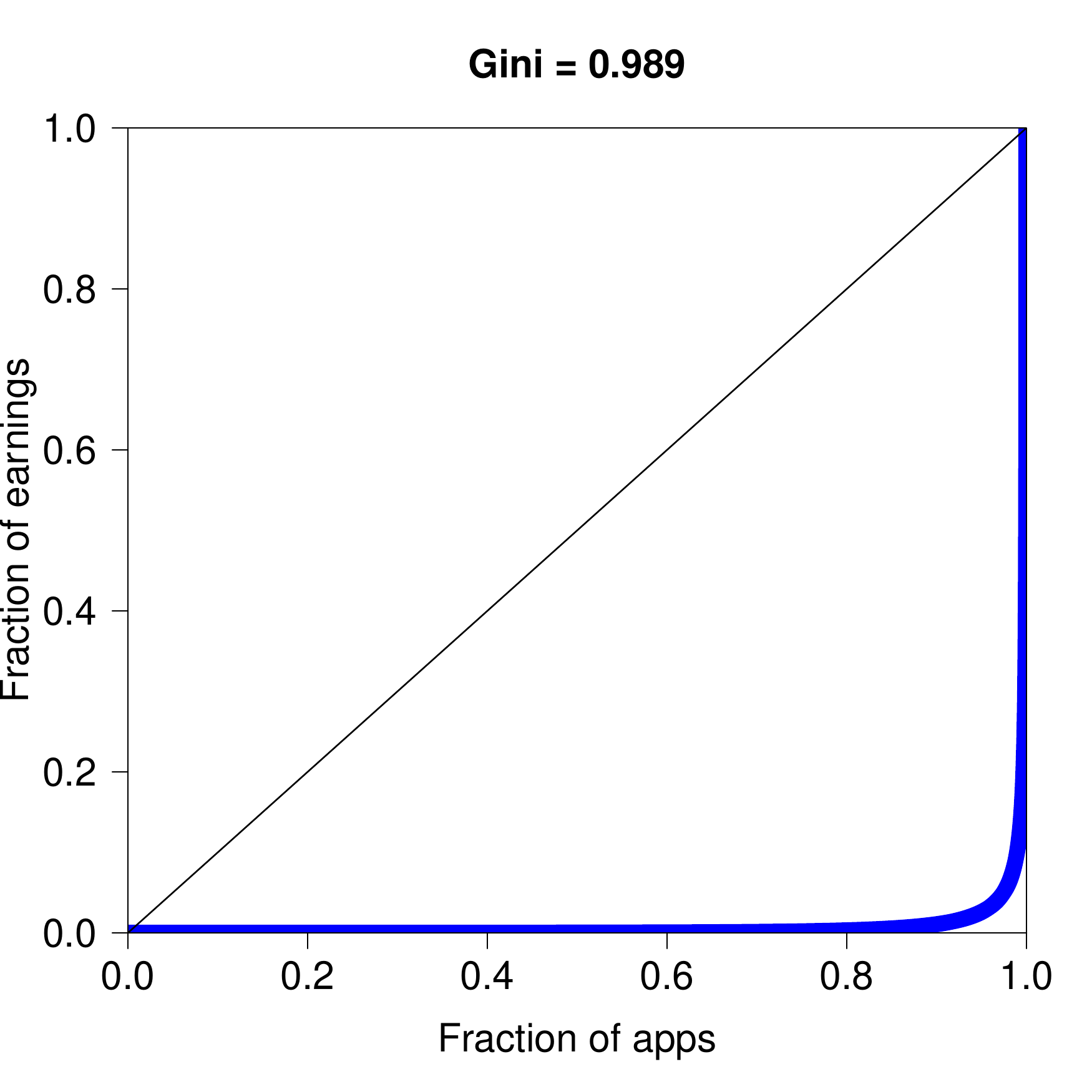}
\end{center}
\vspace*{-1mm}
\caption{Lorenz curve of the earning of the apps, showing extremely high inequality in the earning of the apps.}
\label{fig:lorenz_curve_in_app_earning}
\vspace*{-1mm}
\end{figure}

To better demonstrate the disparities in spending, we plot the Lorenz curve, which shows the percentage of the total spending by different percentiles of the population when ordered by spending (Figure~\ref{fig:lorenz_curve_in_app_spending}).  The diagonal line represents a perfect equality of spending (i.e., if each person spends the same amount). The larger the distance from the diagonal, the larger the inequality in spending. The figure shows very high inequality: the bottom half of buyers spend less than 2\% of the total amount of money, while the top 10\% are responsible for 84\% of all spending. In fact, \emph{just the top 1\% is responsible for 59\% of all the money spent on in-app purchases}. This inequality can be captured by the Gini coefficient that summarizes the distance from equality in a single number, which turns out to be 0.884, representing extremely high inequality. Interestingly, if we consider the earnings of the apps, the inequality is even higher, with Gini coefficient 0.989, and 0.1\% of the apps earning 71\% of all the in-app purchase income (Figure~\ref{fig:lorenz_curve_in_app_earning}). As a comparison, the Gini coefficient for the income of the US population is 0.469, which is the highest among Western industrialized nations (according to census data).

As mentioned above, the top 1\% of buyers, representing 154K users, are responsible for the majority of in-app purchases. In the rest of this section, we focus on this set of users. We call this set of users {\it big spenders}. We also calculated the top 1\% of spenders in each month separately. Among the big spenders, who are the top spenders over the entire 15 month period, 68.4\% are the top 1\% of spenders for half of these months or fewer. This shows that there are bursts and pauses in individual spending levels for each big spender.

\subsection{Characteristics of big spenders}

We start by comparing the demographics of big spenders with the rest of the users in our data set. Understanding the differences in demographics could be useful for advertisers and app stores to better target the population that is more likely to be a big spender.

Big spenders are 22\% relatively more likely to be men (55\% men vs. 45\% women). Regardless of gender, big spenders tend to be older. Men who are big spenders have a median age of 37 years, while the median age in our data set is 34 years. The difference is even larger among women: 43 vs. 35 years. Moreover, there are considerable differences in country of residence statistics for big spenders compared to the typical user. For some countries, like the US, a random user is less likely than average to be a big spender, but for the other countries users are much more likely to be big spenders. For example, Greek, Turkish, and Romanian users are respectively 50, 33, and 29 times more likely to be big spenders than users from our study population at large.

\begin{figure}[t!]
\begin{center}
\includegraphics[width=0.9\columnwidth]{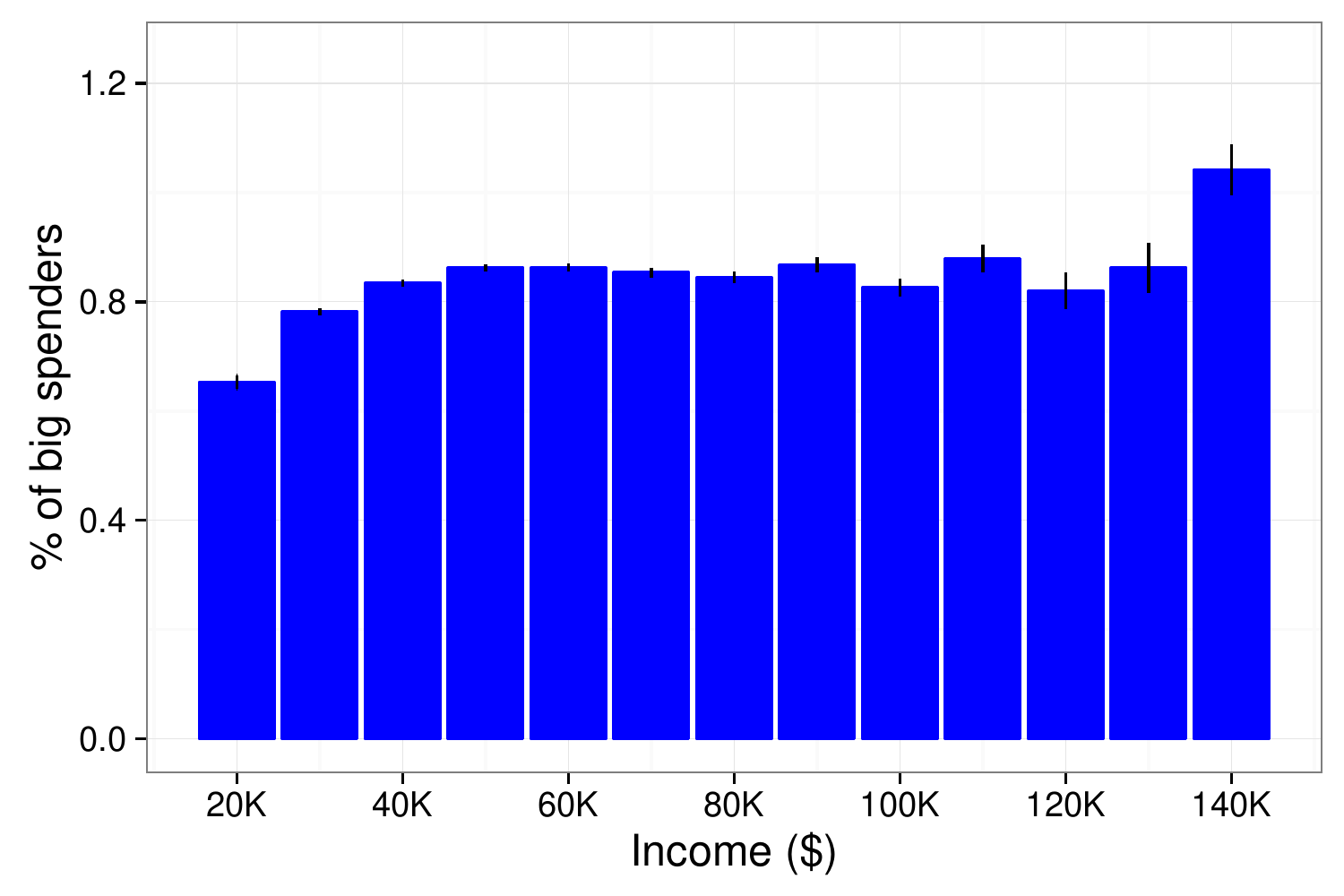}
\end{center}
\caption{Fraction of big spenders, given the income of the users.}
\label{fig:income_avid}
\vspace*{-1mm}
\end{figure}

We also consider the role of income for people from the US, by calculating the fraction of people with a given income who are big spenders. Figure~\ref{fig:income_avid} shows that income has a very small effect on users being big spenders, except for the users with less than \$20K or more than \$140K annual income. Note that the percentage is almost always smaller than the expected 1\%, which is how we define big spenders, because this analysis is conducted only on users from the US, and big spenders are less prevalent in the US.

\subsection{App adoption and abandonment}

To understand app adoption and abandonment, we focus on how users start making in-app purchases within apps, and how they abandon them. In order to analyze the behavior of users over a long period of time within apps they use frequently, we only considered pairs of users and apps that had more than 50 in-app purchases. Furthermore, we were interested in the entire time span of the user's app usage, i.e. from the first time they make a purchase to the last time they make a purchase, so we filtered out the cases in which the usage started before or ended after the period of data collection. This was done by considering only the (user, app) pairs for which the first purchase happened after the first month in our data set, and the last purchase was before the last month.  We call these users {\it frequent buyers}.

\begin{figure}[t!]
\begin{center}
\begin{tabular}{@{}c@{}c@{}}
\subfigure[Change in delay]{
\includegraphics[width=0.49\columnwidth]{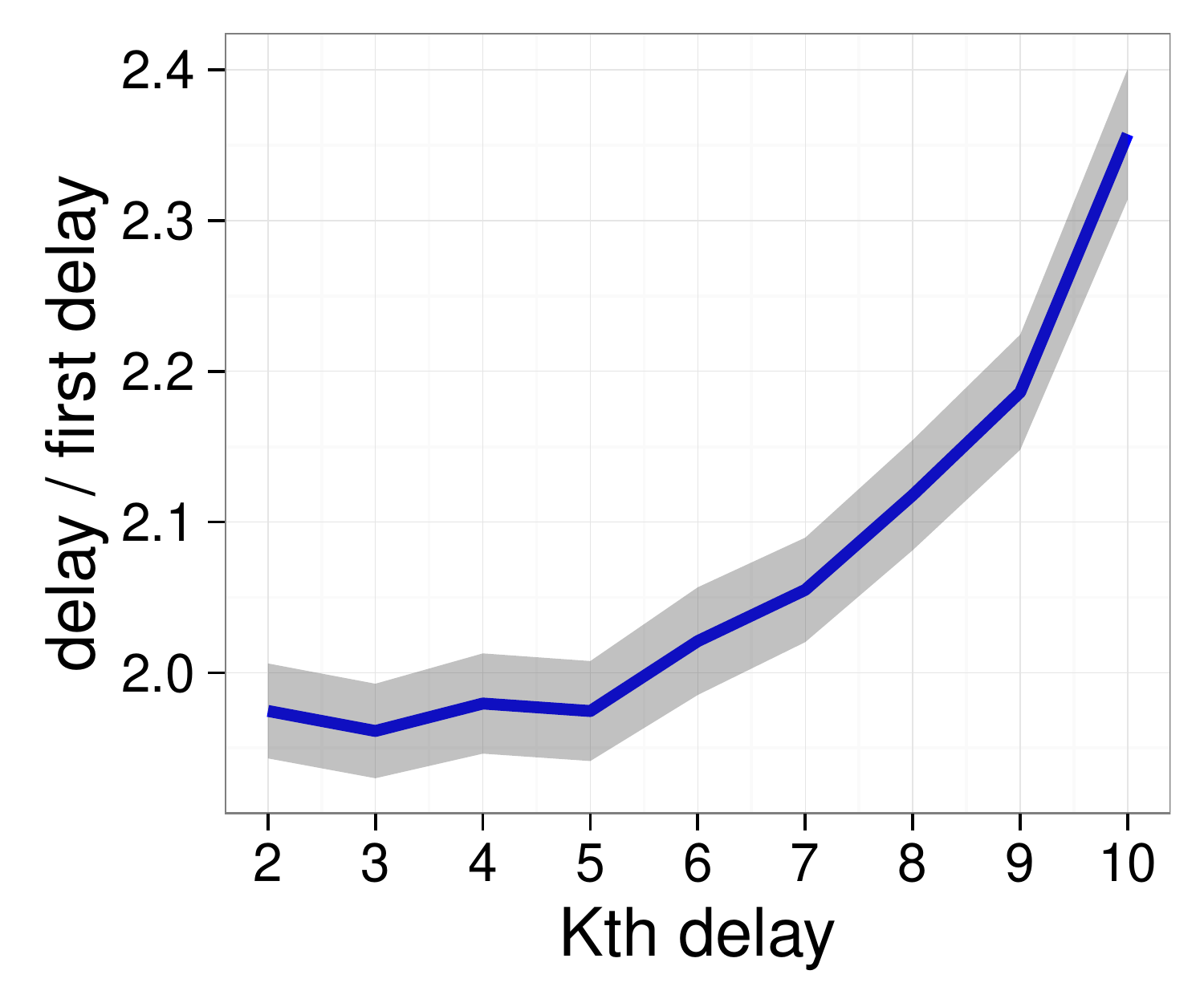}
\label{fig:start_delay_change}
}
&
\subfigure[Change in spending]{
\includegraphics[width=0.49\columnwidth]{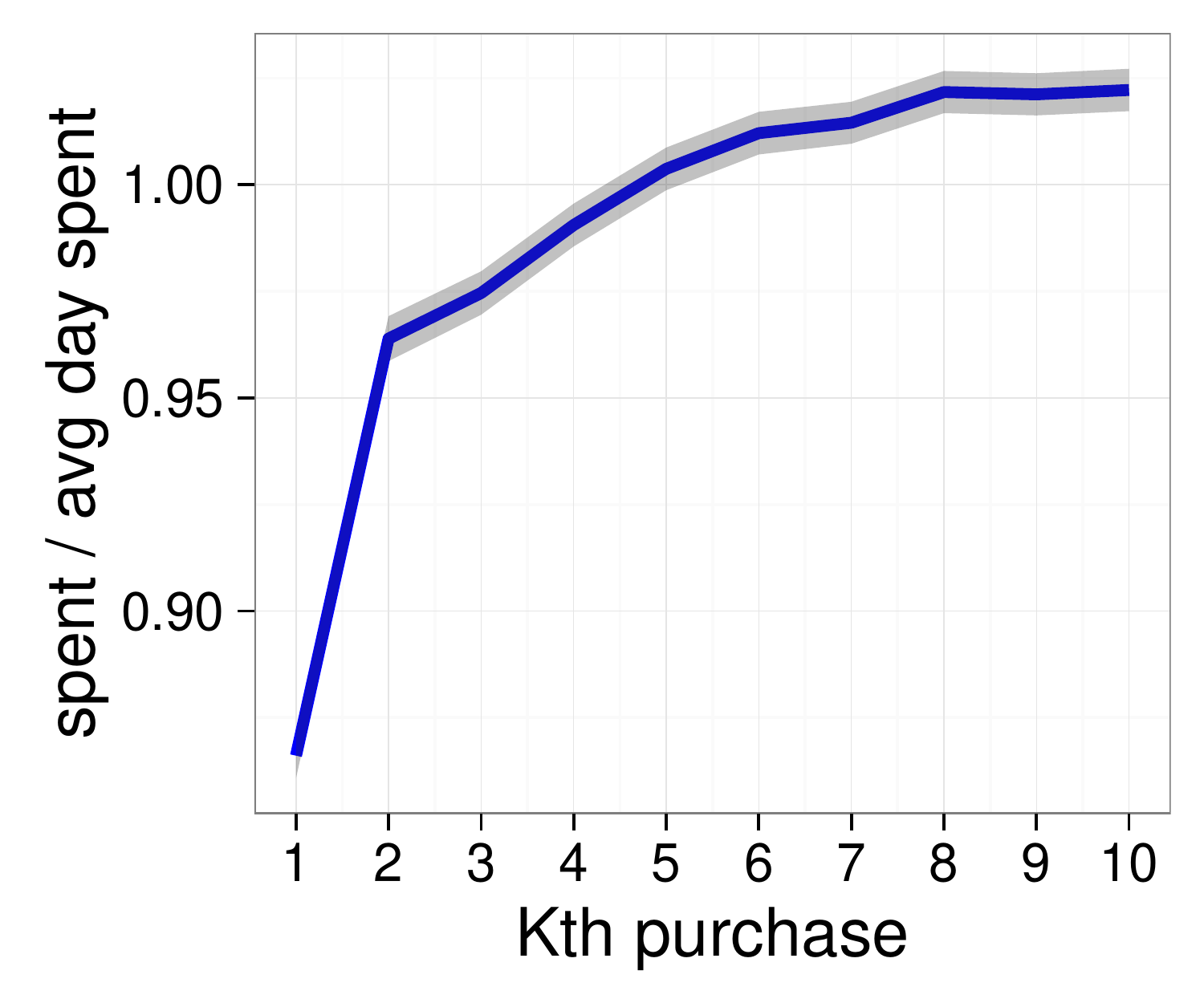}
\label{fig:start_spend_change_avg}
}
\end{tabular}
\vspace*{-2mm}
\caption{Normalized change in time between purchases (i.e. delay) and spending in the first 10 purchases from an app, showing that the delay between consecutive purchases increases, while more money is spent on each purchase.}
\label{fig:adoption_change}
\end{center}
\vspace*{-3mm}
\end{figure}

We start our analysis by looking at the time delay between consecutive purchases. Because our data has one-day granularity, we count multiple purchases by a user in one day as a single, more expensive, purchase. To account for the large heterogeneity in time delays between purchases by different users, we normalize the values for each user individually. Figure~\ref{fig:start_delay_change} shows the 2nd to 9th delays, normalized by the first delay. On average, both the time between purchases and the spending per purchase increase (Figure~\ref{fig:start_spend_change_avg}). This means that even though people make fewer in-app purchases, they spend more money after the first couple of transactions. Since a considerable fraction of the spending is for virtual game coins and bonuses, this suggests that users start by buying small packages of coins and bonuses and move on to the larger ones as they progress in the game. Also, as they buy more and more bonuses and/or coins per purchase, they take longer to replenish their supplies by making a follow-up purchase.

\begin{figure}[t!]
\begin{center}
\begin{tabular}{@{}c@{}c@{}}
\subfigure[Change in delay]{
\includegraphics[width=0.49\columnwidth]{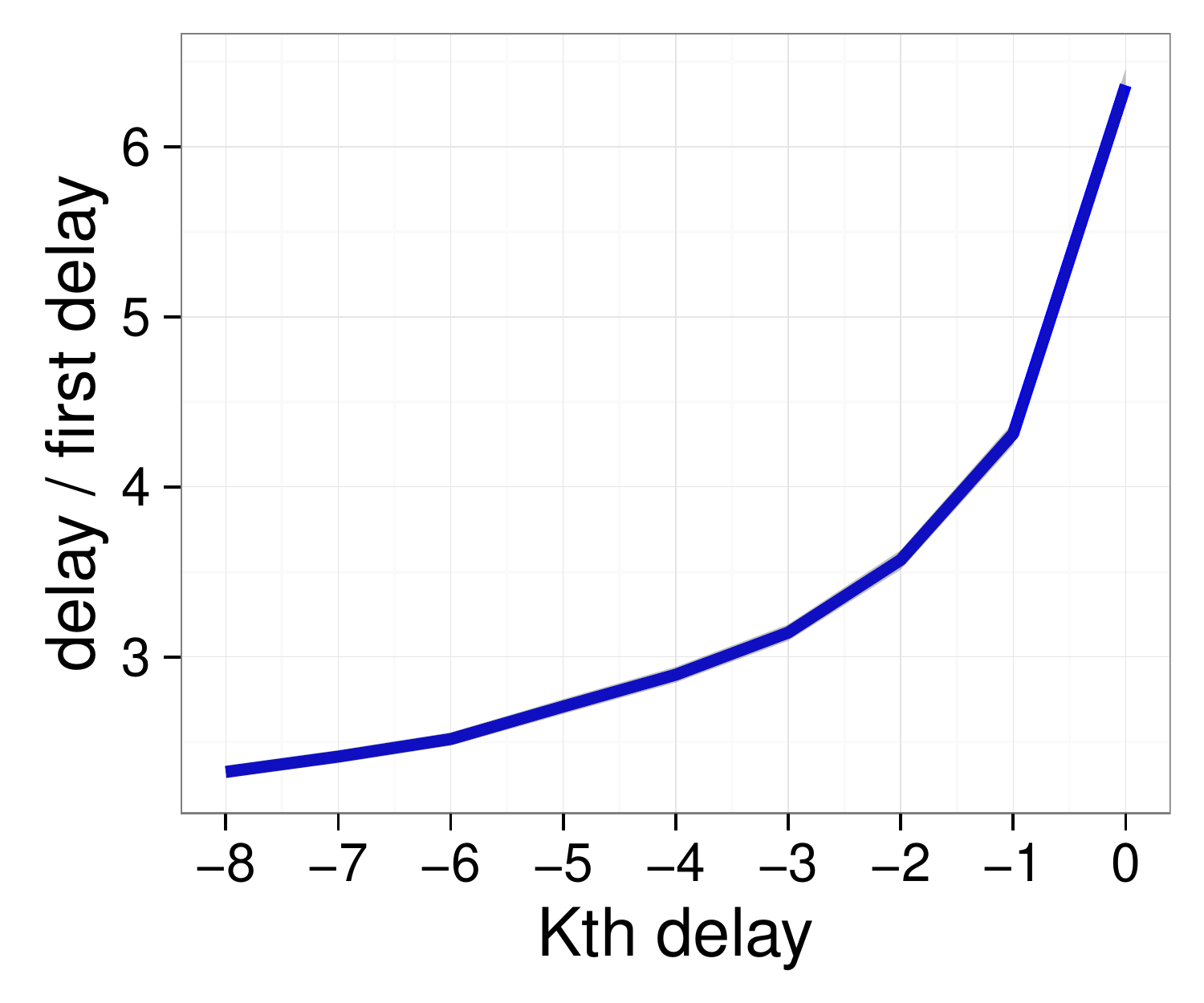}
\label{fig:end_delay_change}
}
&
\subfigure[Change in spending]{
\includegraphics[width=0.49\columnwidth]{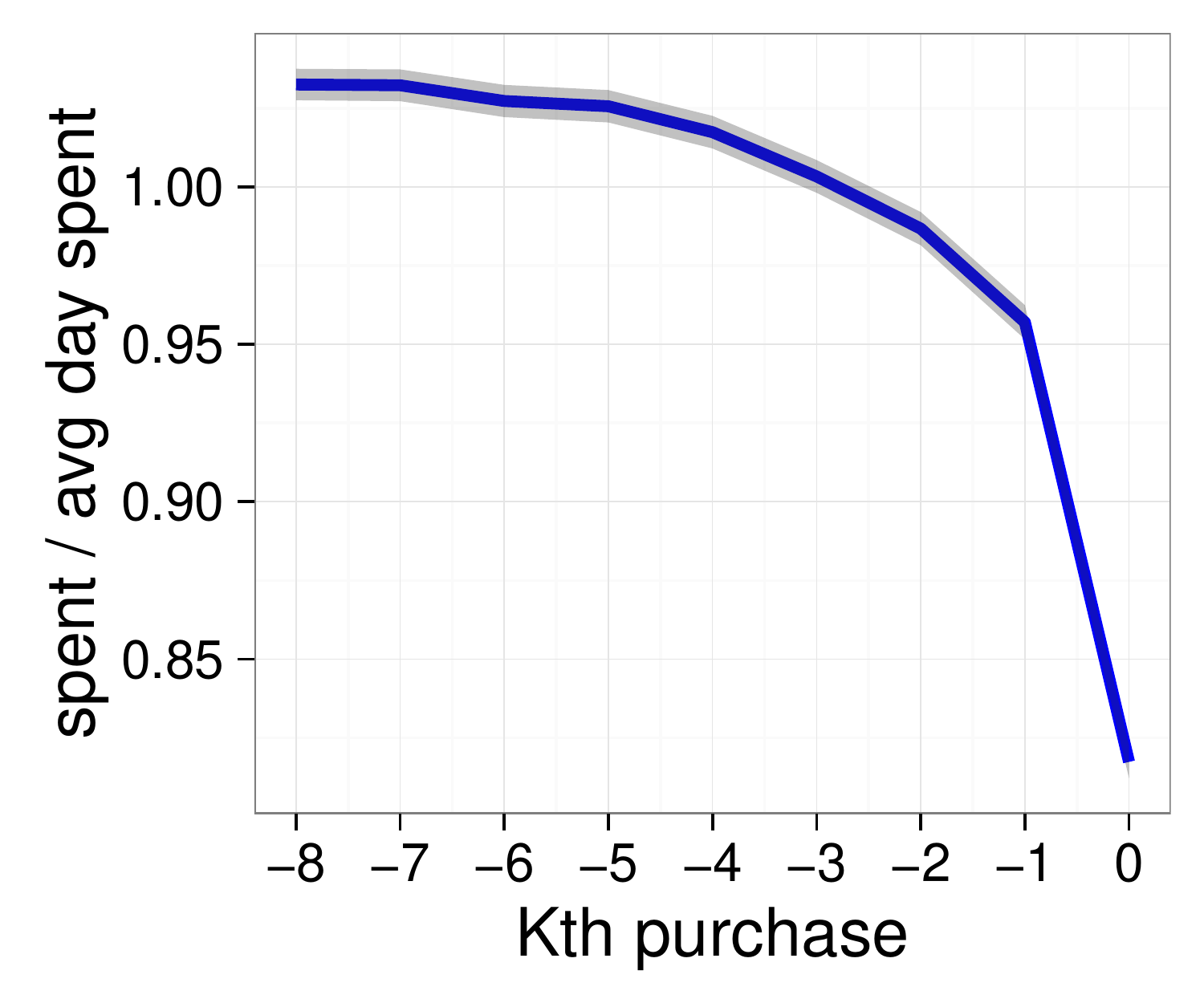}
\label{fig:end_spend_change_avg}
}
\end{tabular}
\vspace*{-2mm}
\caption{Normalized change in delay and spending in the last 10 days with a purchases from an app, showing that the delays become significantly longer and less money is spent on each purchase.}
\label{fig:adandon_change}
\end{center}
\vspace*{-3mm}
\end{figure}

Similarly, when focusing on the user's last 10 purchases within an app, we find that users' delays still get longer, but now at a much higher rate. The very last delay is six times longer than the first delay, on average (Figure~\ref{fig:end_delay_change}). This long delay is a strong indicator of app abandonment. Finally, in Figure~\ref{fig:end_spend_change_avg} we show that as users get closer to their last purchase, they spend less and less money on their daily purchases.

\vspace{4pt}\noindent{\bf Switching to other apps.} Next, we investigate what frequent buyers do after they abandon an app. More precisely, we wanted to find out what fraction of those users switch to another app in which they again become a frequent buyer. We conducted this analysis on the same data as above, i.e., user-app pairs that have more than 50 purchases. We find that $8.6\%$ of frequent buyers who stop making purchases from an app will start making purchases from another app and will become a frequent buyer in the new app (i.e., making more than 50 purchases). This number may seem small, but the frequent buyers who abandon an app are 2.1x more likely to be a frequent buyer in another app compared to a random user from our entire data set (because only $4.1\%$ of users make 50 or more purchases from at least one app). Consequently, from a marketing perspective, it makes much more sense to advertise the new apps to the frequent buyers of existing apps. Furthermore, if we consider a more restrictive definition for frequent buyers and only examine user-app pairs with more than 100 purchases, the difference becomes even larger, and the frequent buyers who have abandoned an app are 4.5x more likely to become a frequent buyer in another app.

%This number may seem small, but if we consider that from all the users in the dataset only 4.1\% are big spenders in at least one app, then the users who abandon an app are 2.1 times more likely than a random user to be a frequent buyer in another app. 

%\note{LA: I don't really understand well the reasoning of the following paragraph. Could you try to rephrase? FK: I explained in a little more details, please let me know which part is not clear.}

\section{Purchase Model}\label{section:model}

\noindent In this section, we model the sequence of purchases people make in order to understand purchasing behavior better. Insights into the future purchasing behavior could be used by both the gaming companies and the app store to increase user engagement and provide better app recommendation.
% KL - why model? In order to predict and explain behavior? FK: Yes, I expanded the first sentence

Following prior work on user consumption sequences \cite{benson2016modeling,anderson2014dynamics}, we model the in-app purchases in 3 main steps: $1)$ modeling time between purchases, $2)$ predicting whether the next in-app purchase will come from an app with a previous purchase by the user, and finally $3)$ predicting the exact app that the user will purchase from, given the output of the previous step. The output of each step is used in the next step; the estimated time interval is one of the main indicators for predicting if the user will purchase from a new app, and we can predict the next app to be consumed much more accurately if we know whether the app is a new app for that user or not.

%Our model has three main steps, similar to Benson et al.~\cite{benson2016modeling}: $1)$ modeling time between purchases $2)$ predicting whether an in-app purchase will be from an app with a previous in-app purchase by the user or a new app for that user, and finally $3)$ predicting the exact app that the user will purchase from, given the output of the previous step. It is possible to model purchases in different order, but we select this sequence to build on the approaches taken in \cite{benson2016modeling,anderson2014dynamics}.

\subsection{Temporal model}

\begin{figure}[t!]
\begin{center}
\begin{tabular}{cc}
\subfigure[Red line shows the theoretical density.]{%
\label{fig:pareto1}
\includegraphics[width=0.45\columnwidth]{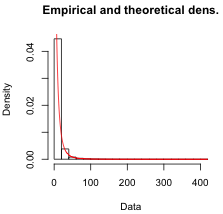}
}
&
\subfigure[Red line represents cumulative distribution function of fitted Pareto distribution.]{%
\label{fig:pareto2}
\includegraphics[width=0.45\columnwidth]{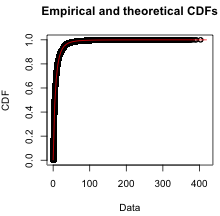}
}
\\
\subfigure[Empirical and theoretical quantiles.]{%
\label{fig:pareto3}
\includegraphics[width=0.45\columnwidth]{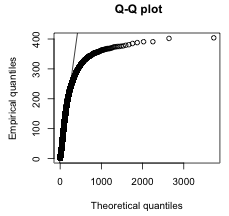}
}
&
\subfigure[Empirical and theoretical percentiles.]{%
\includegraphics[width=0.45\columnwidth]{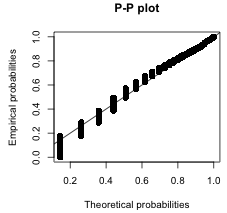}
}
\end{tabular}
\caption{Results of fitting the time between purchases to a Pareto distribution.}
\label{fig:time_dist_fit_pareto}
\end{center}
\end{figure}

First, we investigate a set of parametrized distributions to see which one best describes the distribution of inter-purchase times. We considered Weibull, Gamma, Log normal, and Pareto and find that Pareto best fits the data. We used the Akaike Information Criterion (AIC)~\cite{akaike98information} and the P-P and Q-Q plots, such as the ones shown in Figure~\ref{fig:time_dist_fit_pareto} for Pareto, to compare different distributions. The AIC values were fairly close for all distributions as shown in Table~\ref{table:AIC}, but the plots showed that the Pareto distribution with $shape = 3.21$ and $scale = 20.17$ matches the data better than other distributions. Figures~\ref{fig:pareto1},~\ref{fig:pareto2} show that the modeled distribution fits the probability density function and cumulative density function very well. There are some deviances in the Q-Q plot (Fig~\ref{fig:pareto3}), where the empirical and theoretical quantiles are matched, that shows the distribution failing to capture  very large values in the data. However, the Pareto distribution is still the best fit for our data distribution, considering both the AIC and density distributions.
%\note{should we provide a little bit more detail on the method and on how to interpret the plots?}

\if 0
\begin{figure}[tbh!]
\begin{center}
\includegraphics[width=0.99\columnwidth]{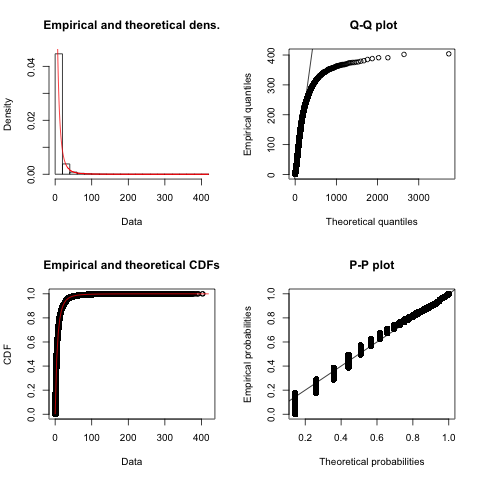}
\end{center}
\caption{Results of fitting the time between purchases to a Pareto distribution.}
\label{fig:time_dist_fit_pareto}
\end{figure}
\fi

\begin{table}[t!]
\caption{AIC for different distributions. Lower AIC scores are preferred.}
\begin {center}
{
\begin {tabular} {l | l}
\specialcellleft{Distribution} & AIC \\
\hline
\hline
Pareto &   59.55M\\
Log Normal &  60.86M\\
Weibull &  61.87M\\
Gamma &  62.22M\\
 \end{tabular}
}
\end{center}
\label{table:AIC}
\vspace*{-6mm}
\end{table}

\subsection{Novelty prediction}\label{subsec:novelty}

Next, we predict whether the user will purchase from a new app or from an app he or she has purchased from in the past. We approach the problem as supervised learning at the time that the user will make the purchase and use the following features: age, gender, time since previous purchase, average time between purchases, average time between re-purchases, total number of purchases, day of the current purchase, percentage of purchases from new vs. past apps, whether the last three purchases are from a new app, and the number of apps from which the user has purchased in the past. We use the first year of our data set for training and the last three months for testing, to avoid using any future information in our predictions.

We tested a collection of different classification algorithms, including several types of decision tree algorithms and SVMs. The C5.0 algorithm in R achieves the best result~\cite{quinlan2004data}. Our classifier achieves a high accuracy, predicting the right class in 84.5\% of cases with precision 0.862, recall 0.965, and F-score 0.964. This accuracy is slightly higher than the result reported in ~\cite{benson2016modeling} for a similar problem on music and video re-consumption.

\begin{figure}[t!]
\begin{center}
\includegraphics[width=0.85\columnwidth]{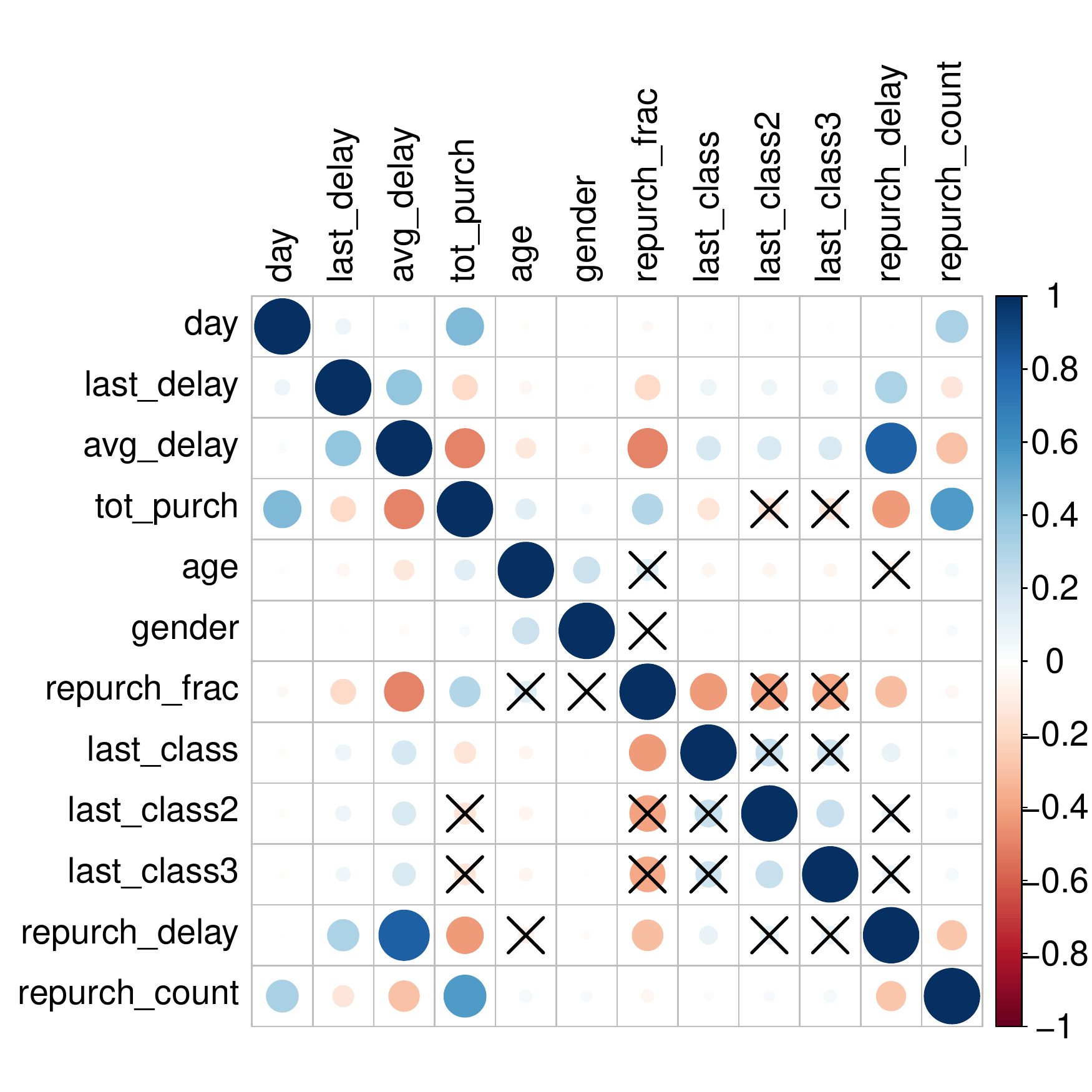}
\end{center}
\vspace*{-2mm}
\caption{Pairwise correlation coefficient among the features for predicting the purchase from new apps. Crossed cells do not have a statistically significant coefficient.}
\vspace*{-2mm}
\label{fig:novelty_corr}
\end{figure}

To better understand the importance of each feature, we also fit a Logistic Regression model to our data set after removing the correlated features. Figure~\ref{fig:novelty_corr} shows the pairwise correlation coefficient between the features. We removed one of the features from pairs with correlation coefficient higher than 0.7. Table~\ref{tab:logistic_reg_novelty} shows the result of the Logistic Regression: the percentage of re-purchases that the user has made is the most important feature; it captures the tendency of the user to re-purchase from an app. The three next most important features capture the user's recent history of re-purchases and purchases from new apps. These are followed by gender, with a positive correlation, showing men are more likely to make a repurchase. This is in tune with our earlier findings that men are more likely to be big spenders, and big spenders make many purchases from the same app.

\begin{table}[t!]
\caption{Results of logistic regression on the independent variables for abandonment prediction. *** $p-value < 0.001$.}
\label{tab:logistic_reg_novelty}
\begin {center}
{
\small{
\begin {tabular} { l | r }
$Variable$ & $Coeff.$ \\
\hline
\hline
\% of re-purchase& $6.236\mathrm{e}{+00}$***\\
Previous class (re-purchase)& $2.878\mathrm{e}{-01}$***\\
2nd to the last class (re-purchase)& $1.624\mathrm{e}{-01}$***\\
3rd to the last class (re-purchase)& $7.878\mathrm{e}{-02}$***\\
Gender (m)& $7.375\mathrm{e}{-02}$***\\
Mean inter-purchase time& $4.764\mathrm{e}{-02}$***\\
Time since last purchase& $-2.782\mathrm{e}{-02}$***\\
Total number of re-purchases& $2.232\mathrm{e}{-02}$***\\
Day of the purchase& $1.236\mathrm{e}{-03}$***\\
Age & $1.069\mathrm{e}{-03}$***\\
\end{tabular}
}
}
\end{center}
\end{table}

\subsection{App prediction}

In two previous steps, we modeled the time between purchases and whether the user's next purchase will come from a new app, with no previous purchases by that user, or from an existing app, with past purchases by that user. If the outcome of the model indicates that the purchase will come from a new app, then our task is to predict the most likely new app, given the previous apps that user made purchases from. If, on the other hand, the outcome of the classifier indicates that the purchase is from an existing app, then we use the sequence of all previous purchases to select the most likely existing app.

%\vspace{4pt}\noindent \textbf{New app prediction.}

\subsubsection{New app prediction}
Given the apps that a user purchased from in the past, our goal is to predict the most likely new app the user will purchase from next. Similar problems have been studied extensively in the area of the recommendation systems~\cite{adomavicius2005toward,bobadilla2013recommender}.

Motivated by the recent success of embedding models in a number of natural language processing tasks~\cite{mikolov2013distributed}, we propose to use a language model to learn app vectors in a low-dimensional space, trained from sequences of user in-app purchases such that apps that appeared in similar context reside nearby in the embedding space. Following the embedding step, we propose to use a k-nearest neighbor approach in the learned vector space to predict the most likely new app given the existing apps consumed by the user.

More formally, let us assume we are given a set of apps $A = \{a_j | j = 1 . . . M\}$, each identified by a unique identifier $a_j$. In addition, in-app purchase times for N users over a time period T from our data set are also known. For the $n^{th}$ user we collect data in a form $d_n = \{(a_j , t_i ), i = 1, . . . , K_i, t_1 < t_2 < . . . < t_{K_{i}}\}$, where $d_n$ denotes the user's in-app purchase sequence, $K_i$ is the total number of in-app purchases user made, and $t_i$ is time of $i^{th}$ in-app purchase from app $a_j$.

\begin{table}[t!]
\begin {center}
\caption{Top 5 closest apps by cosine similarity for 3 apps.}
\label{table:close_apps}
\scalebox{0.95} {
\begin {tabular} {|l|c|}
\hline
 \multicolumn{2}{|c|}{Kim Kardashian West Official}\\ \hline
\hline
 \multicolumn{1}{|c|}{Top 5 closest apps} & Cosine similarity\\
\hline
Khloe Kardashian Official & 0.907\\%495\\
Kourtney Kardashian Official& 0.866\\%857\\
Kylie Jenner Official& 0.863\\%917\\
Kendall Jenner Official& 0.805\\%534\\
kimoji & 0.733\\%694\\
\hline
\multicolumn{2}{|c|}{Homework}\\ \hline
\hline
\multicolumn{1}{|c|}{Top 5 closest apps} & Cosine similarity\\
\hline
Smart Studies& 0.502\\%183\\
iStudy Pro& 0.500\\%390\\
Barrons Hot Words & 0.491\\%156\\
Physics 101 & 0.480\\%380\\
PSAT Preliminary SAT Test Prep & 0.479\\%376\\
\hline
\multicolumn{2}{|c|}{Checkbook Pro}\\ \hline
\hline
\multicolumn{1}{|c|}{Top 5 closest apps} & Cosine similarity\\
\hline
Accounts 2 Checkbook& 0.678\\%532\\
Checkbook Spending & 0.657\\%431\\
Checkbook HD Personal Finance & 0.641\\%007\\
My Check Register& 0.617\\%790\\
My Checkbook& 0.608\\%814\\
\hline
\end{tabular}
}
\end{center}
\end{table}

Given a set $\mathcal{D}$ of $N$ user in-app purchase sequences, where sequence $d_n \in \mathcal{D}$ is defined as an in-app purchase from $K$ apps, the objective is to maximize log-likelihood of the training data $\mathcal{D}$,
\begin{equation}
\mathcal{L} = \frac{1}{N} \sum_{d \in \mathcal{D}} \bigg( \sum_{a_j \in d} \sum_{ -b \le i \ge b, i \ne 0} \log  \P( a_{j+i} | a_j ) \bigg),
\end{equation}
where $b$ is the context widths for in-app purchase sequences and probability $\P( a_{j+i} | a_j )$ of observing a neighboring in-app purchase given the current in-app purchase is defined using a softmax function~\cite{mikolov2013distributed} expressed using app vectors.

Once we learn a vector representation for each app, we can leverage vector cosine similarities to calculate similarities between apps. In Table~\ref{table:close_apps} we show 5-nearest neighbor apps for several randomly selected apps along with the cosine similarities between the corresponding app vectors. As demonstrated in the table, the proposed approach can accurately capture similarities between apps.

We used the first year from our data set to train the app embeddings and leveraged them to predict the new apps users will purchase from in the remaining three months. The prediction used cosine similarity between the apps that the user has already purchased from and all the remaining apps. All predictions were made per user. Specifically, if the user made purchases from $k$ apps in the first year, we used these $k$ apps to predict the new $k/4$ apps the user will likely purchase from in the next three months ($k/4$ was chosen because the test period is one fourth of the training period). The $k/4$ apps were predicted by considering the most similar of each of the $k$ previous apps and selecting a quarter of them randomly.

We predict the exact new app from which the user will make a purchase in 4.7\% of the purchases. This seems very low, but considering that there are more than 154K apps the user can choose from in the context of the recommendation systems, the approach is working considerably well. To further quantify the accuracy of the proposed embedding approach, we compared our recommendation strategy to several baseline models: 1) Non-negative matrix factorization (NMF) approach using the matrix of in-app purchases formed from $\mathcal{D}$; 2) LDA~\cite{lda2003} applied on the app description text; 3) ranking the apps by popularity and always predicting top $k/4$ apps users still have not made a purchase from yet.

The results, presented in Table~\ref{table:baselines_knn}, show that the app embedding approach outperforms the considered baselines. Poor performance of LDA agrees with previous research~\cite{lda2010twitter} that found that this method performs poorly when trained on short text documents. Top apps also achieves a low accuracy, due to users experimenting with many less popular apps. Better performance of app embeddings over the NFM approach can be explained by the fact that the NFM model loses the notion of time and sequence order once it transforms the data set $\mathcal{D}$ into the matrix. We also experimented with other ways of selecting the most similar apps, such as considering larger and smaller numbers of candidates, and in all cases our method outperformed all three other baselines with very similar margins to the one reported.

\begin{table}[t!]
\caption{New app prediction accuracy. }
\begin {center}
{
\begin {tabular} {l | l}
\specialcellleft{Method} & Accuracy \\
\hline
\hline
App embeddings & 4.7\%  \\
NMF & 4.1\% \\
Top apps & 2.2\% \\
LDA & 1.7\% \\
 \end{tabular}
}
\end{center}
\label{table:baselines_knn}
\end{table}

%\vspace{4pt}\noindent \textbf{Existing app in-app re-purchase.}
\subsubsection{Existing app in-app re-purchase}
In the case that the model from Section~\ref{subsec:novelty} predicts that a re-purchase is most likely to happen, we use the frequency and recency of the user's previous app consumption to predict the app from which the re-purchase will occur. This may appear to be an easy prediction problem, as one might think that users almost always purchase from the last app they purchased from, or the app from which they made the majority of purchases. However, in case of in-app re-purchases, only 46.5\% of re-purchases come from the latest app they purchased from, and only 45.3\% of them come from the app from which the user made most of the purchases. This justifies the need for a more involved re-purchase model.

We follow a similar approach to the one from~\cite{benson2016modeling,anderson2014dynamics}, and use both recency and popularity of previous apps to predict from which app the user's next in-app purchase will come. We use a weight function and a time function that maps the frequency of the usage and time since previous usage to the learnt values. This repeat consumption model could be used for the $i$th consumption:

\begin{equation}
\P (x_i = e) = \dfrac{\sum_{j<i}^{} I(x_j=e)s(x_j) T(t_i-t_j)}{\sum_{j<i}^{} s(x_j) T(t_i-t_j)}
\end{equation}

In this equation, function $s$ represents the frequency of the purchase from the app, and function T represents the time between the purchases. These functions are optimized jointly by calculating the negative log-likelihood over the equation. The negative log-likelihood is not convex in $s$ and $T$, but is convex in each function when the other one is fixed. Thus, we use a standard gradient descent to maximize the likelihood with respect to $s$ and $T$, separately. After learning the weight functions, we are able to predict the correct app from which the user is going to make a purchase with 54.8\% accuracy, which is considerably higher than the baselines mentioned above, i.e. 46.5\% and 45.3\% accuracy by always predicting the latest or the most consumed app, respectively.

\section{Related Work}\label{section:related}

% Online shopping: our work and etc.
\noindent  Online shopping is becoming more popular as people learn to trust online payment systems, which was not the case in the past~\cite{bhatnagar00risk}. Multiple studies, aimed at profiling online shoppers, found that online shoppers tend to be younger, wealthier, and more educated compared to the average Internet user~\cite{zaman02internet,swinyard03people,swinyard11activities,farag07shopping}. A more recent work showed that while women are more likely to be online shoppers, men spend more money per purchase and make more purchases overall~\cite{kooti2015portrait}. In our work, we focus on a particular subset of online purchases, iPhone digital purchases. There are considerable differences in characteristics of iPhone purchases and purchases of physical goods. One of the main differences is that people are much more likely to purchase the same item multiple times.

% Spending on phone
Similar to online shopping, spending on mobile digital goods is increasing, and people have spent more than \$20 billion dollars in the Apple App Store in 2015~\cite{app_high_spending}, which is four times more per user than in Android App Stores\footnote{\url{http://fortune.com/2014/06/27/apples-users-spend-4x-as-much-as-googles/}}. This might be due to the different demographics of iPhone users. Given this high level of spending, understanding the market would help us to more effectively target apps toward users who are likely to become regular users and frequent spenders. Despite the popularity of the iPhone digital market, there has not been any large-scale study of how people are spending money on this platform. In this work, we show that most of the money is spent on in-app purchases, and we present a demographic and prediction analysis of spending.

Usage and purchases from apps have been the subject of a few studies. Sifa et al. studied the purchase decisions in free-to-play mobile games~\cite{sifa2015predicting}. They built a classifier that predicts whether a user is going to make any purchase in the future and also built a regression model to estimate the amount of money that will be spent by each user. The models are moderately accurate. Schoger studied the monetization of popular apps in the global market, identifying growing markets and that in-app purchases are increasingly accounting for a larger fraction of total purchases~\cite{schoger2014most}. Our study, unlike those studies, includes the full history of iPhone purchases by the users and considers that many users make purchases from multiple apps. Moreover, the large scale of the data set allows us have enough big spenders to analyze their behavior accurately.

% Attrition, bordem, abondanment
We also study changes in user purchases over time, how users becomes frequent buyers in a particular app, and how their purchases evolve over time. The abandonment of a service is called \textit{consumer attrition} or \textit{churn}. The importance of consumer attrition analysis is driven by the fact that retaining an existing consumer is much less expensive than acquiring a new consumer~\cite{rechinhheld1990zero}. Thus, prediction of consumer churn is of great interest for companies, and has been studied extensively. For example., Ritcher et al. exploit the information from users' social networks to predict consumer churn in mobile networks~\cite{richter2010predicting}. Braun and Schweidel focus on the causes of churn rather than when churn will occur~\cite{braun2011modeling}. They find that a considerable fraction of churn in the service they studied happens due to reasons outside the companies' control, e.g., the consumer moving to another state. In the context of mobile games, Runge et al. study user churn for two mobile games and predict it using various machine learning algorithms~\cite{runge2014churn}. They also implement an A/B test and offer players bonuses before the predicted churn. They find that the bonuses do not result in longer usage or spending by the users.  Kloumann et al. study the usage of apps by users who use a Facebook login and model the lifetime of apps using the popularity and sociality of apps, showing that both of these affect the lifetime of the app~\cite{kloumann2015lifecycles}. Baeza-Yates et al. addressed the problem of predicting the next app the user is going to open through a supervised learning approach~\cite{BaezaYates15predicting}. In our work, we model the whole sequence of purchases that users make, including adoption, churn, and prediction of the next app.

Our work is the first work that studies the details of all iPhone purchases made by a large number of users. This allows us to better understand the interplay between usage of multiple apps that are competing for the same users, their attention and their purchasing power.

\section{Conclusion}\label{section:conclusion}

\noindent Mobile devices have grown wildly in popularity and people are spending more money purchasing digital products on their devices. %Despite the increasing popularity, there have not been large scale studies of people's spending on the phone's digital market. 
To better understand this digital marketplace, we studied a large data set of more than 776M purchases made on iPhones, including songs, apps, and in-app purchases. We find that, surprisingly, 61\% of all the money spent is on in-app purchases, and a small group of users are responsible for most of this spending: the top 1\% of users are responsible for 59\% of all spending on in-app purchases. We characterize these users, showing that they are more likely to be men, older, and less likely to be from the US. Then, we focus on how these big spenders start and stop making purchases from apps, finding that as users gradually lose interest, the delay between purchases increases. The amount of money spent per day on purchases initially increases, then decreases, with a sharp drop before abandonment. Nevertheless, from the perspective of app developers these big spenders are a valuable user segment as they are 4.5x more likely to be a big spender in a new app than a random app user. In the last part of our study, we model the purchasing behavior of users by breaking it down into three different steps. First, we model the time between purchases by testing a variety of different distributions, and we find the Pareto distribution fits the data most accurately. Second, we take a supervised learning approach to predict whether a user is going to make purchase from a new app. Finally, if the purchase is from a new app, we use a novel approach to predict the new app based on the previous in-app purchases. If the purchase is from an app that the user purchased from in the past, we combine the earlier frequency of the purchases and the time between the purchases to predict from which app the re-purchase will come. The models proposed in our study can be leveraged by app developers, app stores and ad networks to better target the apps to users.

\subsubsection*{Acknowledgements}
This work was supported in part by the ARO (W911NF-15-1-0142).

%\footnotesize
\balance
% END OF PAPER - bibliography follows
%\vspace{-1mm}
\bibliographystyle{abbrv}
{
%%\small

}
\end{document}